\documentclass[12pt
]{JHEP3}         
\usepackage{amssymb,amsfonts}
\usepackage{cite}
\usepackage{amsmath}
\usepackage{graphicx}
\usepackage{MnSymbol}

\def\rmd{{\rm d}}

\newcommand{\bs}{\boldsymbol}
\newcommand{\el}[1]{\label{#1}}
\newcommand{\er}[1]{\eqref{#1}}
\newcommand{\df}[1]{\textbf{#1}}

\newcommand{\ke}{\rangle}
\newcommand{\br}{\langle}
\newcommand{\lb}{\left(}
\newcommand{\rb}{\right)}
\newcommand{\lc}{\left.}
\newcommand{\rc}{\right.}
\newcommand{\lsb}{\left[}
\newcommand{\rsb}{\right]}

\newcommand{\nn}{\nonumber \\}
\newcommand{\p}{\partial}

\newcommand{\ba}{\begin{eqnarray}}
\newcommand{\ea}{\end{eqnarray}}
\newcommand{\be}{\begin{equation}}
\newcommand{\ee}{\end{equation}}
\newcommand{\bal}{\begin{align}}
\newcommand{\eal}{\end{align}}
\newcommand{\bay}[1]{\left(\begin{array}{#1}}
\newcommand{\eay}{\end{array}\right)}
\newcommand{\eg}{\textrm{e.g.} }
\newcommand{\ie}{\textrm{i.e.}, }
\newcommand{\iv}[1]{{#1}^{-1}}
\newcommand{\st}[1]{|#1\ke}
\newcommand{\at}[1]{{\Big|}_{#1}}
\newcommand{\zt}[1]{\textrm{#1}}

\newcommand{\Tr}{\mbox{Tr}}
\def\xa{{\alpha}}

\def\xb{{\beta}}

\def\xd{{\delta}}
\def\xD{{\Delta}}
\def\xe{{\epsilon}}

\def\xg{{\gamma}}
\def\xG{{\Gamma}}
\def\xk{{\kappa}}

\def\xl{{\lambda}}
\def\xL{{\Lambda}}
\def\xo{{\omega}}
\def\xO{{\Omega}}
\def\xvp{{\varphi}}
\def\xs{{\sigma}}
\def\xS{{\Sigma}}
\def\xt{{\theta}}

\def\dxa{{\dot \alpha}}
\def\dxb{{\dot \beta}}

\def\CH{{\cal H}}

\def\CM{{\cal M}}
\def\CN{{\cal N}}
\def\CO{{\cal O}}

\def\CV{{\cal V}}

\def\BH{\mathbb{H}}

\def\BR{\mathbb{R}}
\def\BS{\mathbb{S}}

\preprint{}
\title{${\cal N} = 4$ Super-Yang-Mills on Conic Space as Hologram of STU Topological Black Hole}

\author{Xing Huang${}^1$, Yang Zhou${}^{2,3}$ \\

${}^1$Department of Physics, National Taiwan Normal University\\ Taipei, 116, Taiwan\\
${}^2$School of Physics and Astronomy, Tel-Aviv University\\ Ramat-Aviv 69978, Israel\\
${}^3$Department of Physics, Hanyang University\\ Seoul 133-791, Republic of Korea\\
{\tt E-mails :xingavatar@gmail.com, yangzhou1984@gmail.com}
}

\abstract{ We construct four-dimensional ${\cal N}=4$ super-Yang-Mills theories on a conic sphere with various background R-symmetry gauge fields. We study free energy and supersymmetric R\'enyi entropy using heat kernel method as well as localization technique. 
We find that the universal contribution to the partition function in the free field limit is the same as that in the strong coupling limit, which implies that it may be protected by supersymmetry. Based on the fact that, the conic sphere can be conformally mapped to $\mathbb{S}^1\times\mathbb{H}^3$ and  the R-symmetry background fields can be supported by the R-charges of black hole, we propose that the holographic dual of these theories are five-dimensional, supersymmetric STU topological black holes. We demonstrate perfect agreement between ${\cal N}=4$ super-Yang-Mills theories in the planar limit and the STU topological black holes.
}

\begin{document}

\pagestyle{plain} \setcounter{page}{1}
\newcounter{bean}
\baselineskip16pt


\section{Introduction}
The rigid supersymmetry of gauge theories in curved backgrounds allows us to compute exact results of a certain class of BPS observables, for instance, the partition function of four-dimensional ${\cal N}=2$ theories in the Omega background \cite{Nekrasov:2002qd} and on a round four-sphere \cite{Pestun:2007rz}.
The generalizations to other dimensions and also other curved backgrounds have been explored extensively.
A far from complete list of references includes \cite{Kapustin:2009kz, Jafferis:2010un, Hama:2011ea, Hama:2010av, Imamura:2011uw, Imamura:2011wg, Alday:2013lba, Hama:2012bg, Nosaka:2013cpa, Closset:2013sxa, Cassani:2014zwa, Assel:2014paa, Kallen:2012cs, Kim:2012ava, Imamura:2012xg, Imamura:2012bm, Alday:2014rxa, Alday:2014bta, Peelaers:2014ima,Nishioka:2014zpa}. 

The goal of this paper is to extend the previous investigations \cite{Nishioka:2013haa, Huang:2014gca, Nishioka:2014mwa} to the supersymmetric gauge theories on four-dimensional spheres with conical singularities. One of the motivations is to compute exact R\'enyi entropy. For conformal field theories (CFTs), the flat space R\'enyi entropy with spherical entangling surface can be mapped to that on a four-sphere where the entangling surface is mapped to the great two-sphere. 
In three dimensions, the authors of \cite{Nishioka:2013haa} studied $\CN\ge2$ Chern-Simons matter theory
on the $q$-branched three-sphere $\mathbb{S}^3_q$ with certain background vector fields turned on to maintain rigid supersymmetry. They computed the supersymmetric partition function, with which a quantity called supersymmetric R\'enyi entropy (henceforth SRE) is defined
\be
S^{\text{super}}_q={ \log Z_{q}(\mu(q)) - q\log Z_{1}(0) \over 1-q}\ .\label{defSRE}
\ee
It can be considered as the generalization of the usual R\'enyi entropy, which instead
has the non-supersymmetric partition function $Z_{q}(0)$ (with zero chemical potential) in the definition (\ref{defSRE}). On the other hand we know that superconformal theories can also be studied via holography and in particular R\'enyi entropy can be computed from the thermal entropy of topological black hole \cite{mann,brillet,vanzo,roberto,birm}.
It has been shown \cite{Huang:2014gca} that free energy and also the supersymmetric R\'enyi entropy computed from CFT on $\mathbb{S}^3_q$ and those computed holographically from the four-dimensional charged BPS topological black hole (TBH) agree exactly, which motivated the authors to propose TBH$_4$/qSCFT$_3$ correspondence (See also \cite{Nishioka:2014mwa} where Wilson loop was discussed). 

In this paper, we study the correspondence between superconformal field theories on the $q$-branched four-sphere $\mathbb{S}^4_q$ and the five-dimensional charged BPS topological black hole (TBH$_5$/qSCFT$_4$).
The branched sphere $\mathbb{S}^4_q$ is a singular space and generally the conical singularity breaks the supersymmetry globally. To compensate the singularity one can turn on a background Abelian R-symmetry gauge field to provide an extra holonomy (around the singularity) so that some of the Killing spinors survive. For ${\cal N}=4$ SYM, the Cartan subgroup of the SO(6) R-symmetry group is U(1)$^3$, and therefore we have multiple choices for the R-symmetry background fields. 
We consider generic backgrounds with one or more U(1) fields turned on. In each background, free energy and supersymmetric R\'enyi entropy in the zero coupling limit can be computed by heat kernel method, taking into account the holonomy contribution.

To perform localization computation, we consider $\mathbb{S}^4_q$ as the singular ($\xe \to 0$) limit of the smooth resolved space $\mathbb{\widehat S}^4_q(\xe)$ \footnote{It is also possible to cut off the cone at radius $\xe$ and impose boundary conditions on fields. This is the so-called ``Hard Wall" prescription. But here we instead use the ``smooth cone" prescription.}. Following \cite{Hama:2012bg}, we construct the $\CN=2$ supersymmetric gauge theories on the resolved space by working out the background field configuration that admits Killing spinors. In the particular case ${\cal N}=4$ SYM,
the singular limit of this construction returns to the background with two equal U(1) fields. We then use the associated supercharge $\df Q$ (to be precise $\widehat{\df Q} = \df Q + \df Q_B$, where $\df Q_B$ is the BRST operator) to localize the path integral of the partition function to a matrix model. Interestingly, we find that the partition function is insensitive to the resolving function, and therefore is identical to the
one on a squashed four-sphere (ellipsoid) \cite{Hama:2012bg}.

We evaluate the matrix integral of ${\cal N}=4$ SYM
in the supergravity limit (large $N$, large 't Hooft coupling $\xl$). In this
limit, the instanton contribution can be neglected.
We solve the saddle point equation and find that the $q$-dependence of the free energy
(and therefore also the SRE) factorizes. We note that in even dimensions, the universal term is the one with logarithmic divergence.
Even though the free energy we get has been regularized, divergence can be restored by the replacement $\log \xl \to \log \xl - \log \lb \frac \ell\xL \rb^2$. \footnote{See \eg \cite{Russo:2012ay} for the relation between the divergent
and the finite parts of the free energy on a round sphere.}

In the $q \to 1$ limit, the coefficient of the logarithmic term is proportional to the $a$-anomaly (see \eg \cite{Duff:1993wm} and references therein), which is a protected quantity and independent of the coupling constant. It is tempting to guess the same independence may hold for the $q>1$ generalization since we have unbroken supersymmetry.
Extracting the $q$-dependent coefficient of the logarithmic term of free energy we find that it is the same as the one in the zero coupling limit and therefore it is protected.

To find the gravity solution dual to $\CN=4$ SYM on $\mathbb{S}^4_q$ we turn
to the so-called STU model, which is a particular five-dimensional $\CN=2$ gauged supergravity theory with three Abelian gauge fields. We note that the STU model can be embedded \cite{Cvetic:1999xp} into ${\cal N}=8$ $\zt{SO}(6)$ gauged supergravity, which is the known gravity dual of $\CN=4$ SYM on a round sphere. Since the U(1)$^3$ Cartan subgroups on both sides are identical, it is natural to expect that the topological black holes carrying one or more U(1) charges are the duals of $\CN=4$ SYM on $\BS^1\times \BH^3$ ( and therefore on $\mathbb{S}^4_q$ by conformal mapping ) with the corresponding U(1) background gauge fields turned on. We find perfect agreement between the two sides. 
 

The rest of the paper is organized as follows. In Sec~\ref{sec:qSCFT}, we study $\CN=4$ supersymmetric gauge theories on a branched four-sphere $\mathbb{S}^4_q$. We consider various types of background that preserve supersymmetry on $\mathbb{S}^4_q$. In the case with two equal U(1) background gauge fields turned on, we compute the exact partition function using localization technique. We also argue what the partition function should look like in the other cases. We then solve the saddle point equation in the supergravity limit and obtain the $q$-dependence of free energy and supersymmetric R\'enyi entropy. Moreover, we compute the $q$-dependence in the zero coupling limit and find the results are the same as those in the strong coupling limit. In Sec~\ref{sec:TBH}, we study general R-charged BPS topological black holes in the STU model. In Sec~\ref{sec:TBH_qSCFT}, we propose that these black holes can be regarded as the gravity duals of ${\cal N}=4$ SYM on $\mathbb{S}^4_q$. We show that the supersymmetric R\'enyi entropy as well as the free energy obtained from the gravity side precisely agree with the corresponding field theory results. We conclude and discuss future questions in Sec~\ref{sec:conclusion}.

\section{qSCFT$_4$}
\label{sec:qSCFT}
Four-dimensional $\CN=2$ supersymmetric field theories on a round sphere have been studied extensively in~\cite{Pestun:2007rz}. This work has been generalized to squashed four-sphere~\cite{Hama:2012bg}\cite{Nosaka:2013cpa}. A systematic approach to construct supersymmetric theories on four-manifold has been developed in \cite{Festuccia:2011ws,Dumitrescu:2012ha,Dumitrescu:2012at} (see also \cite{Klare:2012gn, Cassani:2012ri, Klare:2013dka}). The basic idea is to take the rigid limit of four-dimensional supergravity that couples to the R-multiplet of the field theory. Along this line, $\CN=1$ supersymmetric theories on complex four-manifolds with various topologies, such as $\mathbb{S}^2\times \mathbb{T}^2$ \cite{Closset:2013sxa} and $\mathbb{S}^1\times \mathbb{S}^3$ \cite{Cassani:2014zwa,Peelaers:2014ima,Assel:2014paa,Nishioka:2014zpa} have been explored.\footnote{See \cite{Romelsberger:2005eg}\cite{Kinney:2005ej} for seminal works on four-dimensional superconformal index.} In this section, we refine the previous constructions to the four-sphere with a conical singularity, denoted by $\mathbb{S}^4_q$. The conical singularity is specified by a real parameter $q\in \mathbb{R}$. We can think of $\mathbb{S}^4_q$ as a $q$ deformation of the round four-sphere $\mathbb{S}^4$. Unlike in three dimensions, in four dimensions the systematic construction of supersymmetric field theories on generic manifolds with sphere topology is still lacking as the aforementioned approach does not apply to this case. Therefore, in what follows, we use a less systematic approach in which we particularly focus on the $q$ deformation of round sphere. Our goal is to explore interesting physical results depending on $q$, such as free energy and supersymmetric R\'enyi entropy. 

We will mainly focus on ${\cal N}=4$ SYM with different types of supersymmetric backgrounds, all of which admit solutions to Killing spinor equations. In one particular case, partition function can be computed exactly using localization technique and the result is valid for general ${\cal N}=2$ theories. In this case we will see that the field theory on $\mathbb{S}^4_q$ shares interesting feature with that on a squashed four-sphere \cite{Hama:2012bg}, although the former was motivated by computing supersymmetric R\'enyi entropy and the latter was motivated by Alday-Gaiotto-Tachikawa (AGT) correspondence. This equivalence is the four-dimensional analogy of the similar relation between three-dimensional ellipsoid and branched sphere. We use some hand-waving arguments to conjecture the form of the partition function in other backgrounds. In the supergravity (large $\xl$ and large $N$) limit the matrix integral of the partition function can be evaluated using saddle point method. Interestingly, in this limit the $q$-dependence of the free energy (and SRE) completely factorizes just like in three-dimensions. We also perform the heat kernel computation in the free field limit and find that the $q$-dependence remains exactly the same.

\subsection{Killing spinors on $\mathbb{S}^4_q$}
\label{subsec:Killingbranched}
As a common knowledge of constructing rigid supersymmetric field theories in curved spacetime, one needs to set up the Killing spinor equations. Those equations will generally tell us what backgrounds allow a set of Killing spinors, which generate rigid supersymmetries. The Killing spinors on a round four-sphere $\mathbb{S}^4$ were well explored in the pioneering work~\cite{Pestun:2007rz}, where the metric was presented as a warped form of the flat metric in $\mathbb{R}^4$. We start with a metric representing $\mathbb{S}^4$ as the blowing up of round three-sphere with manifest $\zt U(1)\times \zt U(1)$ toric structure,\footnote{These coordinates are particularly convenient for later use. Namely it can be easily mapped to a hyperbolic space $\mathbb{S}^1\times \mathbb{H}^3$ by a Weyl rescaling. } whose metric is
\be
\rmd s^2/\ell^2 = \rmd\theta^2 + \sin^2\theta\rmd \tau^2 + \cos^2\theta\rmd\phi^2\ .
\ee
Replacing $\rmd\phi^2$ by a two sphere one obtains
\be\label{4sphere1}
\rmd s^2/\ell^2 = \rmd\theta^2 + \sin^2\theta\rmd \tau^2 + \cos^2\theta ( \rmd\phi^2 + \sin\phi^2\rmd\chi^2)\ ,
\ee where the domains of coordinates are specified by
\be\label{domain1}
\theta\in [0,\pi/2]\ ,\quad \tau\in [0, 2\pi)\ ,\quad \phi\in [0, \pi)\ ,\quad \chi\in [0, 2\pi)\ .
\ee
This metric (\ref{4sphere1}) can also be obtained by embedding the four-sphere into $\mathbb{R}^5$
\be
x_0^2+x_1^2+x_2^2+x_3^2+x_4^2 = \ell^2\ ,
\ee and taking the following polar coordinates
\ba
&& x_0 = \ell\, \cos \theta \cos \phi \ ,\nn
&& x_1 = \ell\, \sin \theta \cos \tau \ , \nn
&& x_2 = \ell\, \sin \theta \sin \tau \ , \nn
&& x_3 = \ell\, \cos \theta \sin \phi \cos\chi \ , \nn
&& x_4 = \ell\, \cos \theta \sin \phi \sin\chi \ .\label{Trans}
\ea
The branched four-sphere $\mathbb{S}^4_q$ can be specified by the deformation of $\mathbb{S}^4$. This can be easily seen by dilating the metric while keeping domains of coordinates (\ref{domain1}) intact. The metric of $\mathbb{S}^4_q$ then turns into
\be\label{4sphere2}
\rmd s^2/\ell^2 = \rmd\theta^2 + q^2\sin^2\theta\rmd \tau^2 + \cos^2\theta ( \rmd\phi^2 + \sin\phi^2\rmd\chi^2)\ .
\ee
This space has a conical singularity at $\theta=0$, but regular everywhere else. It can be regarded as a deviation from $\mathbb{S}^4$ parameterized by $q-1$. Therefore we expect that the Killing spinor equations have minimal deviations from those on round sphere, with an additional background vector field $A_\mu$. In 4-spinor notation, they take the forms of 
\ba
& &D_\mu\zeta = +{1\over 2\ell}\gamma_\mu\zeta^\prime\ ,\label{4spinorkilling1}\\
& &D_\mu\zeta^\prime= -{1\over 2\ell}\gamma_\mu\zeta\ ,\label{4spinorkilling2}
\ea
where the background field $A_\mu$ is included in the covariant derivatives,
\be
D_\mu = \nabla_\mu \pm iA_\mu \ .
\ee Notice that we have not put in the indices for the R-symmetry group, which are necessary for theories with ${\cal N}>1$ supersymmetry. In what follows, we study (\ref{4spinorkilling1})(\ref{4spinorkilling2}) on the branched sphere to determine $A_\mu$, which compensates the conical singularity. 
\subsubsection{solution 1}
For the metric (\ref{4sphere2}), the vielbein can be chosen as
\be\label{vielbein1}
e^1 = \ell \rmd\theta\ ,\quad e^2 = q\ell\sin\theta\rmd\tau\ ,\quad e^3 = \ell\cos\theta\rmd\phi\ ,\quad e^4 = \ell \cos\theta\sin\phi\rmd\chi\ ,
\ee and the non-vanishing spin connections are 
\ba
\omega_\tau^{12} =-\omega_\tau^{21} = -q \cos\theta\ ,\quad\omega_\phi^{13} = -\omega_\phi^{31}=\sin\theta\ ,\nn
\quad \omega_\chi^{14}=-\omega_\chi^{41}=\sin\theta\sin\phi\ ,\quad \omega_\chi^{34}=- \omega_\chi^{43}=\cos\phi\label{spinconnection1}\ .
\ea
We choose the following four-dimensional Euclidean gamma matrices expressed
in terms of Pauli matrices as
\ba\label{gammas}
\gamma_1= \left(
\begin{array}{cc}
 0 & i\tau_1 \\
 -i\tau_1 & 0 \\
\end{array}
\right)\ ,\quad \gamma_2=\left(
\begin{array}{cc}
 0 & i\tau_2 \\
 -i\tau_2 & 0 \\
\end{array}
\right)\ ,\nn
\gamma_3=\left(
\begin{array}{cc}
 0 & i\tau_3 \\
 -i\tau_3 & 0 \\
\end{array}
\right)\ ,\quad \gamma_4=\left(
\begin{array}{cc}
 0 & 1_{2\times 2} \\
  1_{2\times 2}  & 0 \\
\end{array}
\right)\ .
\ea
Imposing $\zeta^\prime=-i\zeta$, we see that the two Killing spinor equations
(\ref{4spinorkilling1})(\ref{4spinorkilling2}) coincide
\be
D_\mu\zeta=-\frac {i}{2\ell}\gamma_\mu\zeta\ .
\ee 
For $q=1$, we find the solution with vanishing background field $A_\mu$
\be\label{4killingsolution1}
\zeta_1 = e^{-\frac i2\gamma_1\theta}e^{-\frac12\gamma_2\gamma_1\tau}e^{-\frac12\gamma_3\gamma_2\gamma_1\phi}e^{-\frac12\gamma_4\gamma_3\chi}\zeta_0\ ,
\ee where $\zeta_0$ is a constant spinor
\be
\zeta_0 = \left(
\begin{array}{c}
 0 \\
 c_2 \\
 0 \\
 c_4 \\
\end{array}
\right)\ .
\ee
Now we look for Killing spinor solutions for $q>1$. The strategy is adding background field $A_\mu$ to keep (\ref{4killingsolution1}) still a solution. We find that, with the background field
\be\label{backgroundvalue1}
A_{\mathbb{S}^4_q} = {q-1\over 2}\rmd\tau\ ,
\ee which couples to Killing spinor through the covariant derivative
\be
D_\mu = \nabla_\mu -i A_\mu\ ,
\ee (\ref{4killingsolution1}) is still a solution for the $q$-branched four-sphere. There is also a Killing spinor that has opposite R-charge and satisfies the Killing spinor equation with $\zeta^\prime = i\zeta$
\be
 (\nabla_\mu +i A_\mu)\, \zeta = \frac {i}{2\ell}\gamma_\mu\zeta\ .
\ee This solution is given by
\be\label{4killingsolution2}
\zeta_2 = e^{\frac i2\gamma_1\theta}e^{-\frac12\gamma_2\gamma_1\tau}e^{-\frac12\gamma_3\gamma_2\gamma_1\phi}e^{-\frac12\gamma_4\gamma_3\chi}\widetilde\zeta_0\ ,
\ee where $\widetilde\zeta_0$ is a constant spinor
\be
\widetilde\zeta_0= \left(
\begin{array}{c}
 c_1 \\
0 \\
 c_3 \\
0 \\
\end{array}
\right)\ .
\ee 
\subsubsection{solution 2}
\label{kssolution2}
One can also consider the round four-sphere as a three-sphere fibered on the $\rho$ direction. The metric is given by
\be\label{metric22}
\rmd s^2/\ell^2 = \rmd\rho^2 + \sin\rho^2 (\rmd\theta^2 + \sin^2\theta\rmd \tau^2 + \cos^2\theta\rmd\phi^2)\ .
\ee The vielbein can be chosen as
\ba
e^1/\ell &=& \sin\rho \sin(\tau+\phi) \rmd\theta + \sin\rho \cos(\tau + \phi) \sin\theta \cos\theta (\rmd\tau -\rmd\phi)\ ,\nn
e^2/\ell &=& -\sin\rho \cos(\tau+\phi) \rmd\theta + \sin\rho \sin(\tau + \phi) \sin\theta \cos\theta (\rmd\tau -\rmd\phi)\ , \nn
e^3/\ell &=& \sin\rho\,(\sin\theta^2\rmd\tau + \cos\theta^2\rmd\phi)\ ,\quad e^4/\ell = \rmd \rho\ .\label{vielbein22}
\ea 
We can define a $T$ matrix as 
\be
T(\rho) = \left(
\begin{array}{cccc}
 0 & 0 & -\frac{1}{2} \tan \frac{\rho }{2} & 0 \\
 0 & 0 & 0 & -\frac{1}{2} \tan \frac{\rho }{2} \\
 \frac{1}{2} \cot \frac{\rho }{2} & 0 & 0 & 0 \\
 0 & \frac{1}{2} \cot \frac{\rho }{2} & 0 & 0 \\
\end{array}
\right)\ .
\ee With the gamma matrices given in (\ref{gammas}), we find the following
matrix identities
\be
{1\over 4}\omega_\mu = \gamma_\mu T(\rho)\ ,\quad \mu = \theta, \tau, \phi
\ee where $\omega_\mu$ are spin connections. This implies that an arbitrary constant 4-spinor $\zeta_0$ satisfies the first three components ($\mu=\theta, \tau, \phi$) of equations (\ref{4spinorkilling1}), provided
\be\label{4spinorkilling11}
{1\over 2\ell}\zeta^\prime := T(\rho) \zeta\ . 
\ee
 There remains an undetermined $\rho$-dependent matrix factor $S(\rho)$ and the Killing spinor solution will be given by
\be
\zeta = S(\rho) \zeta_0\ .
\ee $S(\rho)$ can be determined by studying the $\rho$ component of equation (\ref{4spinorkilling1}) and it is given by
\be
S(\rho)=\left(
\begin{array}{cccc}
 \sin \frac{\rho }{2} & 0 & 0 & 0 \\
 0 & \sin \frac{\rho }{2} & 0 & 0 \\
 0 & 0 & \cos \frac{\rho }{2} & 0 \\
 0 & 0 & 0 & \cos \frac{\rho }{2} \\
\end{array}
\right)\ .
\ee
Now the $q$-branched four-sphere is obtained by simply replacing $\rmd\tau$ in (\ref{metric22})(\ref{vielbein22}) by $q\rmd\tau$, and it is straightforward to see that 
\be\label{zeta3}
\zeta = S(\rho)\left(
\begin{array}{c}
c_1 \\
c_2 \\
c_3 \\
c_4 \\
\end{array}
\right)\ .
\ee
is still a solution, provided that a background field is turned on through the coupling
\be
D_\mu=\nabla_\mu + i A_\mu \left(
\begin{array}{cccc}
 1 & 0 & 0 & 0 \\
 0 & -1 & 0 & 0 \\
 0 & 0 & 1 & 0 \\
 0 & 0 & 0 & -1 \\
\end{array}
\right)\ ,
\ee where $A_\mu$ takes the value
\be
A_{\mathbb{S}^4_q} = {q-1\over 2}\rmd\tau\ .
\ee The solution (\ref{zeta3}) can be decomposed into 2-spinors $(\xi, \bar\xi)$ following Appendix \ref{app:4to2}
\be
\xi = \sin{\rho\over 2} \left(
\begin{array}{c}
 c_1 \\
 c_2 \\
\end{array}
\right)\ ,\quad \bar\xi = \cos{\rho\over2} \left(
\begin{array}{c}
 c_3 \\
 c_4 \\
\end{array}
\right)\ .
\ee
One can further introduce subscript indices $A,B$ ($A,B=1,2$) to denote R-charges, and the solution (\ref{zeta3}) can be decomposed as $\xi_A$ and $\bar\xi_A$
\be\label{roundks}
\xi_1 = \sin{\rho\over2}\left(
\begin{array}{c}
 c_1 \\
 0 \\
\end{array}
\right)\ ,\quad \xi_2 = \sin{\rho\over2}\left(
\begin{array}{c}
0 \\
c_2 \\
\end{array}
\right)\ ,\quad \bar\xi_1 = \cos{\rho\over2}\left(
\begin{array}{c}
 c_3 \\
 0 \\
\end{array}
\right)\ , \quad \bar\xi_2 = \cos{\rho\over2}\left(
\begin{array}{c}
0 \\
c_4 \\
\end{array}
\right)\ .
\ee In terms of the R-charge indices $A,B$ the background field in the 2-spinor notation can be written
in a matrix form
\be\label{RchargeB}
[A_\tau]^A_{\,\,\,\,B} = {q-1\over 2} \left(
\begin{array}{cc}
 1 & 0 \\
 0 & -1 \\
\end{array}
\right)\ .
\ee As we will see later in (\ref{diagonalB}), this background can be embedded into SU(2)$_R$ background as the diagonal part.
\subsection{From CFT on $\mathbb{S}^4_q$ to CFT on $\mathbb{S}^1\times \mathbb{H}^{3}$}
\label{conformal_mapping}
One of the motivations to study the supersymmetric branched sphere is to
compute the supersymmetric R\'enyi entropy \cite{Nishioka:2013haa}. Let us first go over the basic definitions of R\'enyi entropy and its supersymmetric generalization. Consider a quantum state, or more generally a density matrix $\rho$ defined on a spatial slice that consists of two regions $A$ and $B$ separated by the entangling surface $\xS$. We can trace over degrees of freedom in the region $B$ and obtain a reduced density matrix $\rho_A = \Tr_B \rho $. R\'enyi entropy for $\rho_A$ is defined by
\be
S_q = \frac 1 {1-q} \log \Tr(\rho_A^q)\,.
\ee
For QFT, the $q$th-power of density matrix can be expressed in terms of the partition function
\be
\Tr(\rho_A^q) = Z_q/(Z_1)^q,
\ee
where $Z_q$ is the partition function on the $q$-fold cover of the original Euclidean spacetime. The $q \to 1$ limit then gives the entanglement entropy across $\xS$. This method to compute the entanglement entropy is the so-called replica trick. 

Most of the time $Z_q$ is difficult to compute for interacting quantum field theories. However the computation may be greatly simplified when supersymmetry is preserved on the covering space
and localization techniques become available. Generally supersymmetry is broken globally on the covering space and we need to turn on certain background fields in order to have unbroken supercharges. The supersymmetric quantity
to compute is
\be
\el{superRenyi}
S_q = \frac 1 {1-q} \log \lb \frac {Z_q(\mu)} {Z_1(0)^q}\rb,
\ee
where $Z_q(\mu)$ is the partition function on the $q$-fold covering space with nonvanishing background gauge field (or equivalently chemical potential $\mu$). This gauge field couples to the R-current. Note that \er{superRenyi}
is similar yet different from the charged R\'enyi entropy \cite{Belin:2013uta}.
The latter contains $Z_1(\mu)^q$ (instead of $Z_1(0)^q$) in the denominator and therefore (generally) is not a supersymmetric quantity. We notice however that the $q\to 1$ limit in either case gives the entanglement entropy. In the remaining of this paper, we will focus on the SRE on four-sphere, which is related to the flat space $\BR^4$ by a conformal mapping. The entangling surface becomes the great two-sphere under the mapping and the $q$-fold cover is the branched sphere $\mathbb{S}^4_q$.

Other than exploring the possibility of supersymmetric localization, the problem of computing (supersymmetric) R\'enyi entropy can also be approached with the help of conformal mapping. A CFT on $\mathbb{S}^d_q$ can be mapped to that on $\mathbb{S}^1\times \mathbb{H}^{d-1}$ after appropriate Weyl rescaling of the metric. The metric of $\mathbb{S}^d_q$
\be\label{dsphere}
\rmd s^2/\ell^2 = \rmd\theta^2 + q^2\sin^2\theta\rmd \tau^2 + \cos^2\theta \rmd\xS_{d-2,+1}
\ee
can be rewritten under the coordinate transformation
\be
\sinh \eta = -\cot \theta
\ee
in the form
\be
\rmd s^2 = \sin^2\theta\left( \rmd\tau^2 + \ell^2(\rmd\eta^2 +  \sinh^2\eta \rmd\xS_{d-2,+1})\right) \ ,
\ee
where we have defined
\be
\tau = q \tau\ell ,\quad \tau\in [0, 2\pi q\ell)\ ,
\ee and $\rmd\xS_{d-2,+1}$ represents the metric of a unit round $d-2$ sphere.
By dropping the overall Weyl scale factor  $\sin^2\theta$, we get the metric on $\mathbb{S}^1\times \mathbb{H}^{d-1}$
\be\el{StimesH2}
\rmd s^2 = \rmd\tau^2 + \ell^2 (\rmd\eta^2 +  \sinh^2\eta \rmd\xS_{d-2,+1})\ .
\ee
Under the conformal mapping, the North Pole $\theta = 0$ is mapped to the boundary of the hyperbolic space, $\eta \rightarrow - \infty$.

In odd dimensions, the partition functions (whose finite part is physical) of conformal field theories are invariant under the Weyl rescaling
\be
Z[\mathbb{S}^d_q] = Z[\mathbb{S}^1_q\times \mathbb{H}^{d-1}]\ , \quad d~ \text{odd}\ .
\ee This is no longer the case in even dimensions due to conformal anomaly. Yet the coefficient $a$ in front of the log term of partition function
\be
Z[\mathbb{S}^d_q] = \dots\ + a \log\left(\frac \ell\epsilon\right) + \dots\ , \quad d~ \text{even}\ ,
\ee
which is associated with Weyl anomaly and independent of regularization scheme, is expected to be universal and invariant under Weyl rescaling
\be
a[\mathbb{S}^d_q] = a [\mathbb{S}^1_q\times \mathbb{H}^{d-1}]\ , \quad d~ \text{even}\ .
\ee
This allows us to compute the log term of SRE on a sphere by studying the thermal partition function on $\mathbb{S}^1\times \mathbb{H}^{d-1}$. Note that the background gauge field $A$ on $\mathbb{S}^d_q$ is also invariant under the Weyl rescaling since the rescaling only affects the metric.  The computation of non-supersymmetric R\'enyi entropy for a free field theory using this mapping can be found in \cite{Casini:2010kt,Klebanov:2011uf}.
As in the non-supersymmetric case \cite{Casini:2011kv}, the conformal mapping also allows us to identify the SRE of a general CFT on $\mathbb{S}^d$ with the SRE across a spherical entangling surface in $\mathbb{R}^{1,d-1}$. 
In the case of strongly coupled CFTs, this mapping allows one to relate R\'enyi entropy to the thermal entropy of the dual AdS black hole~\cite{Hung:2011nu,Belin:2013uta}. The exact gravity dual of SRE in three dimensions
was found in \cite{Huang:2014gca,Nishioka:2014mwa}.

\subsection{Supersymmetric R\'enyi entropy in free limit}
In this section, we compute the SRE for ${\cal N}=4$ super Yang-Mills on $\mathbb{S}^4_q$ (or a spherical entangling surface in $\mathbb{R}^{1,3}$) in the free field limit. This computation
can be extended to ${\cal N}=2$ and ${\cal N}=1$ conformal field theories straightforwardly. After a conformal mapping
from $\mathbb{S}^4_q$ to $\mathbb{S}^1\times \mathbb{H}^3$, the problem becomes computing the thermal partition function on a hyperbolic space, which can be solved using the heat kernel methods \cite{Casini:2010kt}. Generalization
to the case with nonvanishing gauge field is straightforward~\cite{Belin:2013uta}.

The partition function $Z(\beta)$ on $\mathbb{S}^1_\beta \times \mathbb{H}^d$  can be computed from the heat kernel of the Laplacian operator $\xD$
\be\label{partitionkernel}
\log Z(\beta) = {1\over 2} \int_0^\infty {dt\over t}K_{\mathbb{S}^1\times \mathbb{H}^d}(t)\ ,
\ee where 
\be
K(t):= \Tr (e^{- t \xD}) = \int \rmd^d x\sqrt g K(x,x,t),\quad
K(x,y,t) := \br x|e^{- t \xD}\st{y}\,
,
\ee
and $\beta=2\pi q$ denotes the size of $\mathbb{S}^1$. The heat kernel on a product manifold can be factorized
\be
K_{\mathbb{S}^1\times \mathbb{H}^d}(t)=K_{\mathbb{S}^1}(t)\, K_{\mathbb{H}^d}(t) e^{(d-1)^2\pi^2 t}\ ,
\ee where the exponentiation is to eliminate the gap in the spectrum of the Laplacian on $\mathbb{H}^d$.
The heat kernel on $\mathbb{S}^1$ is known as
\be
K_{\mathbb{S}^1}(t)={\beta\over \sqrt{4\pi t}}\sum_{n\neq 0,\in\mathbb{Z}} e^{-\beta^2n^2\over 4t} \ .
\ee The hyperbolic space $\mathbb{H}^3$ is homogeneous and therefore the volume $V$ factorizes
\be
K_{\mathbb{H}^3}(t)= \int d^3x \sqrt{g}~K_{\mathbb{H}^3}(x,x,t):=VK_{\mathbb{H}^3}(0,t)\ .
\ee The equal-point heat kernel on $\mathbb{H}^{3}$ for a complex scalar is known as
\be
K^b_{\mathbb{H}^3} (0,t)= {2\over (4\pi t)^{d/2}}e^{-(d-1)^2\pi^2 t}\ ,\quad d=3\ ,
\ee while for a Weyl spinor the heat kernel is
\be\label{fermionkernel}
K^f_{\mathbb{H}^3}(0,t) = {2(1+\tfrac t 2)\over (4\pi t)^{d/2}}e^{-(d-1)^2\pi^2 t}\ ,\quad d=3\ .
\ee
Turning on a constant background field 
\be 
A_\tau=\mu/q
\ee
along $\mathbb{S}^1$ gives the heat kernel a phase shift. Making use of the formulae above (\ref{partitionkernel})-(\ref{fermionkernel}), the free energy for a complex scalar on $\mathbb{S}^1_\beta \times \mathbb{H}^3$ can be computed
\be
F^b(\beta,\mu):=-\log Z^b(\beta,\mu)=-V\sum_{n\neq 0,\in\mathbb{Z}}  {1\over 2}\int_0^\infty\left[ {\rmd t\over t}{\beta\over \sqrt{4\pi t}}e^{-n^2\beta^2\over 4t}{2\over (4\pi t)^{3/2}}\right] e^{i2n\pi\mu}\, .
\ee
The free energy for a Weyl spinor can be obtained similarly \footnote{There is an additional overall minus sign compared to scalar.}
\be
F^f(\beta,\mu)=V\sum_{n\neq 0,\in\mathbb{Z}}  {1\over 2}\int_0^\infty\left[ {\rmd t\over t}{\beta\over \sqrt{4\pi t}}e^{-n^2\beta^2\over 4t}{2(1+\tfrac
t 2)\over (4\pi t)^{3/2}}\right] e^{i(2\pi\mu-\pi)n}\, ,
\ee
where we have imposed anti-periodic boundary condition for the spinor at $\mu = 0$. Evaluating $F^b$ and $F^f$ explicitly, we get
\be
F_q^b(\mu):=F^b(2\pi q,\mu)=\frac{V\left(\mu ^4+2 \mu ^3+\mu ^2-\frac{1}{30}\right) }{12 \pi  q^3}\ ,
\ee
and
\be
F_q^f(\mu)=-V\lsb\frac{ 240 \mu ^4-120 \mu ^2+\left(30-360 \mu ^2\right) q^2+7}{2880 \pi  q^3}\rsb\ .
\ee
For fixed $\mu$, one can compute the charged R\'enyi entropy for both scalar and spinor using
\be
S^{\text{charged}}_q={qF_1(\mu)-F_q(\mu)\over 1-q}\ ,
\ee while for SRE, $\mu$ is required to be a function of $q$ with the constraint $\mu(q=1)=0$ because of supersymmetry and therefore
\be
S^{\text{super}}_q={qF_{1}(0)-F_{q}(\mu(q))\over 1-q}\ .
\ee One can easily see that when the field is neutral $\mu=0$, the charged and supersymmetric R\'enyi entropies reduce to the non-supersymmetric one 
\be S^{\text{charged}}_q=S^{\text{super}}_q=S^{\text{non-SUSY}}_q\ .
\ee As a consistent check, one can reproduce the known result of non-supersymmetric R\'enyi entropy for free ${\cal N}=4$ super Yang-Mills (including $6$ real scalars, $4$ Weyl spinors, $1$ vector)~\cite{Fursaev:2012mp} \footnote{Note that we temporarily drop the overall group factor for the theory with SU($N$) gauge group, which is not relevant for the $q$-scaling behavior. This group factor needs to be recovered when we compare the free field results with the localization results as well as the gravity results later.}
\be
S^{\text{non-SUSY}}_q = 6\times{S^b\over 2}+4\times S^f+S^v = \frac{\left(1+q+7q^2 +15q^3\right) V}{48 \pi q^3}\ ,
\ee 
where we have inserted the R\'enyi entropy for a vector field
\be
S^v = \frac{\left(91 q^3+31 q^2+q+1\right) V}{360 \pi  q^3}\ .
\ee
In the rest of the text, we will mainly focus on the SRE and for simplicity of notation we denote it by $S_q$. 

Now we are ready to compute the SRE of ${\cal N}=4$ super Yang-Mills theory. For convenience, we extract the extra contribution in the SRE for each field
$
\Delta S: =S_q - S^{\text{non-SUSY}}_q
$ due to the non-vanishing $\mu$. For a complex scalar
\be
\Delta S^b(\mu)= \frac{(\mu +1)^2 \mu ^2 V}{12 \pi  (q-1) q^3}\ ,
\ee while for a Weyl spinor
\be\label{fermionRenyiE}
\Delta S^f(\mu) = \frac{\mu ^2 \left(-2 \mu ^2+3 q^2+1\right)V}{24 \pi  (q-1) q^3}\ .
\ee
Note that the vector field is neutral under the R-symmetry group. 

As discussed in section \ref{subsec:Killingbranched},
the conical singularity can be compensated by the background gauge fields so that some of supercharges are preserved. These background gauge fields couple to $\zt U(1)$ R-currents. In the case of ${\cal N}=4$ SYM, there are three independent $\zt U(1)$'s as the Cartan subgroup of $\zt{SO}(6)$ R-symmetry. We denote the three $\zt U(1)$'s by $\zt U(1)_i$ and the corresponding background fields by $A^i$ ( chemical potential $\mu_i$ is defined by $A_\tau^i$ and we will omit the subscript $\tau$ from now on). The charges $(k_1, k_2, k_3)$ of the field components of ${\cal N}=4$ multiplet are listed in Table \ref{3charges}.
\begin{table}[htdp]
\caption{charges under three $\zt U(1)$'s}
\begin{center}
\begin{tabular}{l*{7}{c}r}
              &$\psi^1$ & $\psi^2$ & $\psi^3$ & $\psi^4$ & ${\cal A}_\mu$ & $\phi^1$ & $\phi^2$&$\phi^3$ \\
\hline
$k_1$        & $+\frac12$ & $-\frac12$   & $-\frac12$  & $+\frac12$  &0&+1&0&0  \\
$k_2$        &  $-\frac12$ & $+\frac12$   & $-\frac12$  & $+\frac12$ &0&0&+1&0 \\
$k_3$        & $-\frac12$ & $-\frac12$   & $+\frac12$  & $+\frac12$  &0&0&0&+1  \\
\end{tabular}
\end{center}
\label{3charges}
\end{table}%

Because the complex scalars $\phi^i$ and Weyl spinors $\psi^{1,2,3,4}$ of
the ${\cal N}=4$ SYM couple to a few different background gauge fields, we need to determine the effective chemical potential $\mu$, which follows from the weighted (by charges) sum of individual chemical potentials $\mu = k_i \mu_i$. Note that Killing spinors should couple to all background fields $A^i$, although we did not distinguish different background fields when we were solving Killing spinor equations. The charges of the chiral Killing spinors are given in Table~\ref{Killing3charges} 
\footnote{The values in the table follows from a similar table in \cite{Pestun:2007rz}, where the R-symmetry group has been reduced to $\zt{SU}(2)_L^R \times \zt{SU}(2)_R^R \times \zt{SO}(1,1)^R$. In our case, the internal space is no longer Lorentzian and we have the Euclidean version SO(2)$^R$ instead, which can be chosen as U(1)$_3$. We note that four of the six real scalars are charged under the Cartan subgroup of SU(2)$_L^R$ with each pair having the same charge. So this Cartan subgroup is generated by the sum
of the two generators of $\zt U(1)_1 \times \zt U(1)_2$, while the Cartan of SU(2)$_R^R$ is associated with the difference of the two.}, 
\begin{table}[htdp]
\caption{charges of Killing spinors}
\begin{center}
\begin{tabular}{l*{5}{c}r}
             & $\zt{SU}(2)_L $ & $\zt{SU}(2)_R $ & $k_1$ & $k_2$ & $k_3$ \\
\hline
$\xi_A$        & $\df 2$ & $0$   & $\pm\frac12$  & $\pm\frac12$  & $+\frac 1 2$  \\
$\bar \chi_{\dot A}$        &  $0$ & $\df 2$   & $\pm\frac12$  & $\mp\frac12$ & $+\frac 1 2$ \\
$\bar \xi_A$        & $0$ & $\df 2$   & $\pm\frac12$  & $\pm\frac12$  & $-\frac 1 2$  \\
$\chi_{\dot A}$        & $\df 2$ & $0$   & $\pm\frac12$  & $\mp\frac12$ & $-\frac 1 2$ \end{tabular}
\end{center}
\label{Killing3charges}
\end{table}%
where $\zt{SU}(2)_L \times \zt{SU}(2)_R$ is the local rotation group on $\BS^4$.
As we can see, $\xi_A$ and $\bar \xi_A$ are chiral and anti-chiral components of a Dirac spinor. We will discuss various cases in which some of the Killing spinors survive and they are classified according to how many background gauge fields are turned on.

\subsubsection{A Single U(1)}
\label{1chargeB}
We first consider the case with a single background field ($\mu_3 \ne 0$). 
The compensation by gauge field is measured by $k_i A^i$ (or equivalently $k_i \mu_i$). 
Since the chiral Killing spinors have charges $|k_3| = \frac12$, the chemical potential can be determined from the value of the background field (\ref{backgroundvalue1}) 
\be\label{chemical1}
\mu_3 = q-1\ .
\ee
From Table \ref{3charges} we see that there are two pairs of Weyl
fermions charged $\pm\frac12$ respectively and one complex scalar charged $+1$. Note that the contribution to SRE
from fermions (\ref{fermionRenyiE}) is an even function of the chemical potential. The SRE is computed by
\be
S_q = S^{\text{non-SUSY}}_q + 4 \Delta S_f(\mu=\frac{q-1}{2}) + \Delta S_b(\mu=q-1)\ ,
\ee and finally we obtain
\be
{S_q\over S_1}  = 1\ .\label{freeSq1}
\ee
Note that the ratio we discuss here is also the ratio of the universal terms since the common factor, the volume $V$ contains log divergence. 

\subsubsection{Two U(1)'s}
\label{2chargeB}
Next we consider the case with two background fields of equal values ($\mu_1 = \mu_2 \ne 0$). The compensation is given by $\mu_1 (k_1 + k_2)$ and we
have the effective charge given by $r=k_1+k_2$. To make the charged ( $|k_1 + k_2| = 1$ ) Killing spinors still satisfy equations (\ref{4spinorkilling1})(\ref{4spinorkilling2}), the values of chemical potentials should be
\be\label{chemical2}
\mu_1 = \mu_2 = \frac{q-1}{2}\ .
\ee
From Table \ref{3charges} we see that there are two Weyl fermions charged $\pm1$ respectively and two complex scalars charged $+1$. The SRE reads
\be
S_q = S^{\text{non-SUSY}}_q + 2 \Delta S_f(\mu=\frac{q-1}{2}) + 2\Delta S_b(\mu=\frac{q-1}{2})\ ,
\ee and finally we obtain
\be
{S_q\over S_1}  = {3q+1\over 4q}\ .\label{Sq2charges}
\ee
\subsubsection{Three U(1)'s}
\label{3chargeB}
Finally we consider the generic case with all three background U(1) fields turned
on. For the same reason in the two cases above, we can preserve the Killing spinors of equivalent charge $|k_1 + k_2 + k_3| = \frac 32 $ with the choice of chemical potentials
\be\label{chemical4}
\mu_1 = (q-1)\frac{a}{3}\ ,\quad \mu_2 = (q-1)\frac{b}{3}\ ,\quad \mu_3 = (q-1)\lb 1-\frac{a+b}{3}\rb\ .
\ee
We can define the effective charge $r$ for all the charged fields 
\be\el{effrcharge} \lb \frac {q-1} 2\rb r = k_i \mu_i\,.\ee 
From Table \ref{3charges} we see that, effective charges $r$ of the four Weyl spinors are $+1$, $-1+\frac{2 a }{3}$, $-1+\frac{2 b }{3}$ and $-1+\frac{2 a + 2b }{3}$. The three complex scalars are effectively charged $+\frac{2 a }{3}$, $+\frac{2 b }{3}$ and $2-\frac{2 a +2b}{3}$. The SRE is then given by
\ba
S_q &= & S^{\text{non-SUSY}}_q +\Delta S_f(\mu=\frac{q-1}{2}) + \Delta S_f(\mu=\frac{(3-2 a)(q-1)}{6}) \nn &+& \Delta S_f(\mu=\frac{(3-2 b)(q-1)}{6}) +\Delta S_f(\mu=\frac{(3-2 b-2a)(q-1)}{6})\nn 
&+& \Delta S_b(\mu=\frac{(q-1)a}{3}) + \Delta S_b(\mu=\frac{(q-1)b}{3}) \nn &+& \Delta S_b(\mu=\frac 13 (q-1)(3-a-b))\ ,
\ea
and the $q$-dependence is
\be
\label{Sq3charges}
{S_q\over S_1}  = \frac 1 {27 q^2}\left(q^2 C_2+q C_1+C_0 \right)\ ,
\ee with the coefficients
\ba
C_2 &=& -a^2 (-3 + b) - a (-3 + b)^2 + 3 (9 - 3 b + b^2)\ ,\nn
C_1&=&a^2 (2 b-3)+a \left(2 b^2-9 b+9\right)-3 (b-3) b\ ,\nn
C_0&=&-a  b  (a +b -3)\ .
\ea
In the special case with all chemical potentials being equal ($a=b=1$),
\be\label{chemical3}
\mu_1 = \mu_2 =\mu_3 = \frac{q-1}{3}\ ,
\ee
the SRE is computed by
\be
S_q = S^{\text{non-SUSY}}_q + 3\Delta S_f(\mu=\frac{q-1}{6}) +\Delta S_f(\mu=\frac{q-1}{2})  + 3\Delta S_b(\mu=\frac{q-1}{3})\ ,
\ee and the ratio (\ref{Sq3charges}) becomes
\be
{S_q\over S_1}  = \frac{19 q^2+7q+1}{27 q^2}\ .
\ee

\subsection{Exact partition function on $\mathbb{S}^4_q$}
\label{sec:Localization}
In this section the exact partition function of ${\cal N}=4$ super Yang-Mills on the branched four-sphere is studied. In order to do this, we first construct ${\cal N}=2$ SCFT on a resolved branched sphere and then compute its partition function using localization technique. It turns out that the partition function on the branched sphere with background (\ref{chemical2}) and the one on an ellipsoid \cite{Hama:2012bg} are equal, as in the three-dimensional case \cite{Nishioka:2013haa}. We also comment on partition functions in the generic backgrounds (\ref{chemical4}). Finally we study the large $N$ matrix models in the special case of ${\cal N}=4$ SYM on the branched sphere with different types of backgrounds and work out the $q$-dependence of their partition functions and SREs.

\subsubsection{Supersymmetric resolved branched four-sphere}
We recall that the branched four-sphere $\mathbb{S}^4_q$ \er{4sphere2} has a conical singularity at $\xt = 0$. As a common recipe \cite{Fursaev:1995ef} to handle the singularity, one may instead study a sequence of smooth resolved spaces $\mathbb{\widehat S}^4_q(\xe)$ ($\epsilon>0$ is small) and consider $\mathbb{S}^4_q$ as the $\xe \to 0$ limit of $\mathbb{\widehat S}^4_q(\xe)$.
In order to see how the resolving is introduced, we first turn
to the 4d ellipsoid, which is defined by the embedding equation in $\mathbb{R}^5$ ($b:= (\tilde \ell/\ell)^{1/2}$),
\begin{equation}
{x_0^2\over \ell^2}+{x_1^2+x_2^2\over \tilde \ell^2}
 +\frac{x_3^2+x_4^2}{\ell^2} = 1\,.
\label{ellipsoids}
\end{equation}
In particular, for $\tilde\ell=q\ell$, the metric of the ellipsoid is obtained using (\ref{Trans}) (with $\ell\to q\ell$ for $x_1, x_2$),
\be
\el{ellipsoid}
\rmd s^2= f(\xt)^2\, \rmd \xt^2+\ell^2(q^2 \sin ^2\theta\, \rmd\tau^2 + \cos^2\xt(\rmd\phi^2+\sin ^2\phi\, \rmd\chi^2))\ ,\ee
where $f(\theta) = \sqrt{\ell^2(\sin^2\theta+q^2\cos^2\theta)}$. The difference between the singular metric \er{4sphere2} and the smooth one \er{ellipsoid}, implies that we should resolve
the singular metric by adding a factor $f_\epsilon(\theta)$. Thus the metric of $\mathbb{\widehat S}^4_q(\xe)$ is given by
\be
\el{rbranchedsph}
\rmd s^2= f_\epsilon(\xt)^2\, \rmd \xt^2+\ell^2(q^2 \sin ^2\theta\, \rmd\tau^2 + \cos^2\xt(\rmd\phi^2+\sin ^2\phi\, \rmd\chi^2))\ ,\ee
where $f_{\epsilon}\left(\theta\right)$ is a smooth function satisfying
\be
\el{resolvedf}
f_{\epsilon}\left(\theta\right)=\begin{cases}
~q\ell \ , & \theta\rightarrow0 \ \\
~\ell\ , & \epsilon<\theta\leq  \frac{\pi}{2} \ .
\end{cases}
\ee
As we shall see, with appropriate background fields turned on, the resolved space \er{rbranchedsph} allows Killing spinors. For later convenience, from now on we switch to the coordinates $(\rho, \eta, \tau, \chi)$, in which $\tau,\chi$ remains intact but $\rho, \eta$ are related to $\xt,\phi$
by the transformations
\ba
\el{coordtrf}
\sin\theta &=& \sin\eta  \sin\rho\ ,\nn
\tan\phi &=& \cos\eta  \tan\rho\ .
\ea
The metric then becomes
\be
\rmd s^2 = \ell^2\sin^2 \rho(q^2\sin ^2 \eta\rmd
\tau^2+\cos^2\eta  \rmd\chi^2) + (F \sin \rho \rmd \eta + H \rmd \rho)^2
+ G^2 \rmd \rho^2\ ,\label{rbranchedsph1}
\ee
where $F,G,H$ are functions of $\eta, \rho$. Their explicit forms, together with vielbein and spin connection are given in Appendix~\ref{App:Resolved}. Now we study the Killing spinor equations on the resolved branched sphere \er{rbranchedsph1}. The strategy is to require the Killing spinor on a round sphere to remain a solution on the resolved space and we search for the appropriate background configuration for that to happen. Then the Killing spinor equations can be turned into a set of linear algebraic equations of the background fields which have nontrivial solutions.

Following the setup in \cite{Hama:2012bg}, we shall construct ${\cal N}=2$ theories with R-symmetry group $\zt{SU}(2)_R\times \zt U(1)_R$ on the resolved branched sphere (\ref{rbranchedsph1}). Particularly we use a non-Abelian background SU(2)$_R$ gauge field and 2-rank tensor fields $T^{ab}, {\bar T}^{ab}$ to compensate the deviation from the round sphere. The Killing spinor equations consist of {\it main equation} and {\it auxiliary equation}. The former set is essentially extended \er{4spinorkilling1} in the 2-spinor notation \footnote{We use the same notations as that used in~\cite{Hama:2012bg}, see Appendix \ref{app:notations}. The decomposition of 4-spinor to 2-spinor is shown in Appendix~\ref{app:4to2}.} 
\begin{eqnarray}
 D_\mu \xi_A+T^{ab}\sigma_{ab}\sigma_\mu \bar\xi_A &=&
 -i\sigma_\mu \bar\xi'_A\ ,
 \nonumber \\
 D_\mu \bar\xi_A+\bar T^{ab}\bar\sigma_{ab}\bar\sigma_\mu \xi_A &=&
 -i\bar\sigma_\mu \xi'_A\ ,
\label{ks1}
\end{eqnarray}
where $T^{ab}, \bar T^{ab}$ are self-dual and anti-self-dual real background tensor fields, respectively. The covariant derivatives $D_\mu $ are defined with background SU(2)$_\text{R}$ gauge field ${V_\mu }^A_{~B}$ in addition to the spin connection $\Omega_\mu ^{ab}$ \footnote{In this subsection, vielbein and spin connection we take are shown in Appendix \ref{App:Resolved}, which are different from those in Section \ref{kssolution2}.},
\begin{eqnarray}
 D_\mu \xi_A&\equiv& \partial_\mu \xi_A+\frac14\Omega_\mu ^{ab}\sigma_{ab}\xi_A
 +i\xi_B{V_\mu }^B_{~A}\ ,
 \nonumber \\
 D_\mu \bar\xi_A&\equiv& \partial_\mu \bar\xi_A
 +\frac14\Omega_\mu ^{ab}\bar\sigma_{ab}\bar\xi_A +i\bar\xi_B{V_\mu }^B_{~A}\ .
\end{eqnarray}
The set of {\it auxiliary equation}, which follows from extended \er{4spinorkilling2}
in 2-spinor notation reads
\begin{eqnarray}
 \sigma^\mu \bar\sigma^\nu D_\mu D_\nu \xi_A
 +4D_\mu T_{ab}\sigma^{ab}\sigma^\mu\bar\xi_A
 &=& M\xi_A\ ,\nonumber \\
 \bar\sigma^\mu \sigma^\nu D_\mu D_\nu \bar\xi_A
 +4D_\mu\bar T_{ab}\bar\sigma^{ab}\bar\sigma^\mu\xi_A
 &=& M\bar\xi_A\ ,
\label{ks2}
\end{eqnarray}
where $M$ is a background scalar field. 

We choose the particular Killing spinors (\ref{roundks}) on round sphere $\mathbb{S}^4$, which was studied in Section \ref{subsec:Killingbranched}. They can also be presented as 
\begin{eqnarray} \xi_A
 ~=~\left(\xi_1,\xi_2\right)
 &=&\sin\frac\rho2
  \left(\kappa_{_{++}},\kappa_{_{--}}\right),
 \nonumber \\
 \bar\xi_A
 ~=~\left(\bar\xi_1,\bar\xi_2\right)
 &=&\cos\frac\rho2
  \left(i\kappa_{_{++}},-i\kappa_{_{--}}\right),
\label{Kspinors}
\end{eqnarray}
where $\xk_{st}$ are Killing spinors on $\BS^3$ (for $m$ over the coordinates
$(\eta,\tau,\chi)$ and $k,l=1,2,3$),
\begin{equation}
 \Big(\partial_m +\frac14\Omega_m ^{kl}\tau^{kl}\Big)\kappa_{st}=
 -\frac{ist}{2\ell}e^k_m \tau^k\kappa_{st}\ ,\quad
 \kappa_{st}\equiv\frac12\left(\begin{array}{r}
   e^{\frac i2(s\tau+t\chi-st\eta)} \\
 -se^{\frac i2(s\tau+t\chi+st\eta)}
 \end{array}\right)\, .
\end{equation}
Substituting this solution into the auxiliary equation (\ref{ks2}) with vanishing background vector and tensor fields ${V_\mu }^A_{~B}=0,\,T_{ab}=\bar T_{ab}=0$, we get $M=-{1\over 3}R$, where $R$ is the Ricci scalar of $\mathbb{S}^4$.

Now we determine the background fields on the resolved sphere.
First we can regard $\xi_A$ and $\bar\xi_A$ as $2\times 2$ matrices $\bs\xi$ with spinor row indices and SU(2)$_{\rm R}$ column indices. From now on, we use boldface letters to denote $2\times2$ matrix
quantities. In addition to $\bs \xi$, others are 
\begin{equation}
 {\bf V} +V^{[3]}\tau^3\equiv\tilde{\bf V}= E^a\tilde{\bf V}_a\ ,\quad
 i{\bf T} ~\equiv~ \sigma_{ab}T^{ab}\ ,\quad
 i\bar{\bf T} ~\equiv~ \bar\sigma_{ab}\bar T^{ab}\ ,
\end{equation}
and
\begin{equation}
 \bs\xi'= {\bf S}\bs\xi= -i\sigma_{ab}S^{ab}\bs\xi\ ,\quad
 \bar{\bs\xi}'= \bar{\bf S}\bar{\bs\xi}=
 -i\bar\sigma_{ab}\bar S^{ab}\bar{\bs\xi}\ ,
\end{equation}
where $S_{ab},\bar S_{ab}$ are anti-symmetric tensors. In defining
$\tilde{\bf V}$, we subtract the background field $-V^{[3]}\tau^3$ in three dimensions. Note that the spinors $\kappa_{st}$ remain Killing spinors on a three-dimensional resolved branched sphere when $V^{[3]}$  is turned on,
\begin{eqnarray}
 \Big(\partial_m +\frac14\Omega^{kl}_m \tau^{kl}\mp iV_m ^{[3]}\Big)
 \kappa_{_{\pm\pm}}~=~ -\frac{i}{2f_\epsilon}e^k_m \tau^k\kappa_{_{\pm\pm}},
 \nonumber \\
\el{bkg3d} V^{[3]}~\equiv~
 \frac 12\Big(1-\frac{\ell}{f_\epsilon}\Big)\rmd\chi
+\frac 12\Big(1-\frac{q\ell}{f_\epsilon}\Big)\rmd\tau\ .
\end{eqnarray} 
As we will see in the singular limit $\xe \to 0$, all other fields vanish
and $V^{[3]}$ is the only nontrivial background field on the branched four-sphere.
Requiring \er{Kspinors} to remain a solution of the main equation
(\ref{ks1}), we obtain a set of linear algebraic equations for the unknowns
${\bf\tilde V}, {\bf T},{\bf\bar T},{\bf S}$ and $\bar{\bf S}$. In terms of the boldface
notation, they read ($a=4$)
\begin{eqnarray}
  \bs\xi\tilde{\bf V}_4
 +{\bf T}\bar{\bs\xi}
 +\bar{\bf S}\bar{\bs\xi}
 &=&
 i\,\frac{\cos\rho+1}{2G\sin\rho}\bs\xi
 -\frac{H}{2FG\sin\rho}\tau^3\bs\xi
 +\frac{1}{2}\xO_4^{34}\tau^3\bs\xi\ ,
 \nonumber \\
  \bar{\bs\xi}\tilde{\bf V}_4
 +\bar{\bf T}\bs\xi
 +{\bf S}\bs\xi
 &=&
 i\,\frac{\cos\rho-1}{2G\sin\rho}\bar{\bs\xi}
 -\frac{H}{2FG\sin\rho}\tau^3\bar{\bs\xi}
 -\frac{1}{2}\xO_4^{34}\tau^3\bar{\bs\xi}\ ,
\label{kseq2-1}
\end{eqnarray}
and ($a,b=1,2,3$)
\begin{eqnarray}
 \bs\xi\tilde{\bf V}_a
 -i{\bf T}\tau^a\bar{\bs\xi}
 -i\tau^a\bar{\bf S}\bar{\bs\xi}
 &=&
  \frac1{2F\sin\rho}\tau^a\bs\xi
 +\frac12\Omega_a^{b4}\tau^b\bs\xi\ ,
 \nonumber \\
 \bar{\bs\xi}\tilde{\bf V}_a
 +i\bar{\bf T}\tau^a\bs\xi
 +i\tau^a{\bf S}\bs\xi
 &=&
  \frac1{2F\sin\rho}\tau^a\bar{\bs\xi}
 -\frac12\Omega_a^{b4}\tau^b\bar{\bs\xi}\ ,
\label{kseq2-2}
\end{eqnarray} 
where $\xO_a^{b4} := E_a^\mu \xO_\mu^{b4}$ and they take the following form,
\begin{eqnarray}
&&
 \Omega_1^{13}=-\frac 1 F\csc\rho\tan\eta\ ,\quad
 \Omega_2^{23}=\frac{q }F\csc\rho\cot\eta\ ,\nonumber \\
&& 
 \Omega_1^{14}=\frac{ \csc\rho (\cos \rho\,  F+\tan \eta\, H)}{F G}\ ,\quad
 \Omega_2^{24}=\frac{\csc\rho( \cos \rho\, F-\cot \eta \, H)}{F\, G}\ ,\\
&& \Omega_3^{34}=\frac{\p_\rho F+\cot \rho\, F- \csc\rho\,\p_\eta H}{F\,G}\ ,\quad
\xO_4^{34} = -\frac{\csc\rho\, \p_\eta G}{F\, G}\ . \nonumber
\end{eqnarray}
To solve the equations it is helpful to rewrite the action of R-gauge field $\tilde{\bf V}$ on $\bs \xi$ as Gamma matrices acting from the left. This can be done using
\begin{equation}
 \tau^1_\eta\bs\xi=-\bs\xi\tau^3,
\end{equation}
where
\begin{equation}
 \tau^1_\eta~\equiv \tau^1\cos\eta+\tau^2\sin\eta\ ,
\end{equation}
and
\be
 \tau^3\bs\xi=\bs\xi
 \big\{ \cos(\chi+\tau)\tau^1+\sin(\chi+\tau)\tau^2\big\}\ .
\ee
Moreover, we can also express $\bar {\bs \xi}$ in terms of $\bs \xi$ using
\be
 \tau^1_\eta\bs\xi=i\tan\frac\rho2\bar{\bs\xi}\ .
\ee
With all these replacements, every equation in (\ref{kseq2-1}) and (\ref{kseq2-2}) is of the form of a matrix (linear combination of $1$ and $\tau^k$) multiplying $\bs \xi$. The supersymmetric background admitting Killing spinor can be determined by requiring all the matrices to be zero. Note that the manifold with unspecified $F, H, G$ is a generalization of the ellipsoid in~\cite{Hama:2012bg} and the equations for background fields are similar though not the same. 

We nevertheless found nontrivial solutions. The solutions are not unique and can be shifted by solutions to the homogeneous equation, namely the equations (\ref{kseq2-1}) and (\ref{kseq2-2}) with the r.h.s. set to zero. The homogeneous equations are insensitive to the metric and remain the same as those in~\cite{Hama:2012bg}. 
With a properly chosen homogeneous solution, a simple special solution to (\ref{kseq2-1})(\ref{kseq2-2}) is given by ($\tau^2_\eta\equiv i\tau^1_\eta\tau^3$)
\ba
&&{\bf T}  =  \frac14\Big(\frac1F-\frac1G\Big)\tau^1_\eta
 +\frac H{4FG}\tau^2_\eta\ , \quad 
 \bar{\bf T}  =  \frac14\Big(\frac1F-\frac1G\Big)\tau^1_\eta
 -\frac H{4FG}\tau^2_\eta\ , \nn 
&&  {\bf S}  =  -\frac14\Big(\frac1F+\frac1G\Big)\tau^1_\eta
 -\frac H{4FG}\tau^2_\eta\ , \quad 
 \bar{\bf S}  =  -\frac14\Big(\frac1F+\frac1G\Big)\tau^1_\eta
 +\frac H{4FG}\tau^2_\eta\ ,\label{TSbackground}
 \ea
and
 \ba
 && \bs\xi\tilde{\bf V}_1  =
  \frac{\cos \eta \csc \rho\, (G - F)- \sin \eta  \cot \rho H}{2 F G}\tau^1_\eta\bs\xi -\frac{\sin \eta \left[\cot \rho (F-G)+\csc \rho \tan \eta\,  H \right]}{2 F G} \tau^2_\eta\bs\xi\ ,
 \nn 
&&  \bs\xi\tilde{\bf V}_2  =
  \frac{\sin \eta \csc\rho\, (G -F)+ \cos \eta \cot \rho\, H}{2 F G}\tau^1_\eta\bs\xi
 +\frac{\cos \eta \left[\cot \rho (F-G)+\csc \rho \tan \eta\,  H \right]}{2 F G} \tau^2_\eta\bs\xi\ ,
 \nn 
&&  \bs\xi\tilde{\bf V}_3  =
 \frac{\Omega_3^{34}\,F-\cot \rho }{2 F}\tau^3\bs\xi\ ,
 \quad 
  \bs\xi\tilde{\bf V}_4  = 
  \frac{\Omega_4^{34}\, F\, G+\cot \rho\, H}{2 F G}\tau^3\bs\xi\ .
\label{solsp}
\ea
Note that ${\bf T},{\bf\bar T},{\bf S}$ and $\bar{\bf S}$ can be obtained
from the solution on ellipsoid~\cite{Hama:2012bg} by replacing the variables $f,g,h$ (whose explicit forms can be found in \er{ellipfgh}) by $F,G,H$. However, that is not the case for the background gauge field $\tilde{\bf V}$. On the other hand, when $f_\xe (\theta)$ is chosen to be $\sqrt{\ell^2(\sin^2\theta+q^2\cos^2\theta)}$, the background becomes that of ellipsoid.
 
The remaining background scalar field $M$ can be determined straightforwardly.
In $2\times2$ matrix notations, the auxiliary equation (\ref{ks2}) becomes
\begin{eqnarray}
 -4\cot\frac\rho2\Big(\sigma^\mu D_\mu \bar{\bf S}-D_\mu {\bf T}\sigma^\mu \Big)
 \tau^1_\eta
 -4\sigma^\mu \bar{\bf S}\bar{\bf T}\bar\sigma_\mu 
 \nonumber \\ ~=~
 4\tan\frac\rho2\Big(\bar\sigma^\mu D_\mu {\bf S}-D_\mu \bar{\bf T}\bar\sigma^\mu \Big)
 \tau^1_\eta
 -4\bar\sigma^\mu {\bf S}{\bf T}\sigma_\mu 
 &=& M\cdot{\bf 1}\ .
\end{eqnarray}
Plugging in the special solution above (\ref{TSbackground})(\ref{solsp}) we can see that terms with derivatives on $F,G,H$ all cancel and $M$ is given
by
\begin{equation}
 M ~=~ \frac1{F^2}-\frac1{G^2}+\frac{H^2}{F^2G^2}-\frac4{F G}\ .
\end{equation}
{\bf Branched sphere limit}~~
In the singular limit $\xe \to 0$, we get to the branched sphere which has 
\be
F = G = \ell\ ,\quad H = 0\ .
\ee One can immediately see that, in this limit all the fields in (\ref{TSbackground})(\ref{solsp}) vanish except for ${\bf S}$ and $\bar{\bf S}$. \footnote{${\bf S}$ and $\bar{\bf S}$ return to their values on a round sphere.} The only nontrivial background gauge field is
\be\label{diagonalB}
{V_\tau}^A_{~B} = -V_\tau^{[3]}\tau^3 = A_\tau^{U(1)_J}\left(
\begin{array}{cc}
 1 &  \,0 \\
 0 & -1 \\
\end{array}
\right)\ ,\quad A_\tau^{U(1)_J}=\frac {q-1} 2\ ,
\ee which is exactly the background we worked out before (\ref{RchargeB}). We use U(1)$_J$ to denote the Cartan subgroup of SU(2)$_R$.

In the case of ${\cal N}=4$ SYM in the supersymmetric background (\ref{TSbackground})(\ref{solsp}), we shall identify the singular limit as the theory on the branched sphere with two equal U(1) chemical potentials turned on. The latter has been discussed in Section \ref{2chargeB} in details. Gauge fields of $\CN=2$ R-symmetry subgroups $\zt U(1)_J$ and $\zt U(1)_R$ are linear combinations of $A^i$ since both can be embedded in U(1)$^3$. The coefficients of $A^i$ can be obtained by tracing the transformation properties of the scalars. The $\CN=4$ SYM consists of one $\CN=2$ vector multiplet  and one $\CN=2$ hypermultiplet. Each of the three complex scalars represents one of $\zt U(1)_i$ $(i=1,2,3)$, with the charges listed in Table~\ref{3charges}. Following the conventions in~\cite{Hama:2012bg}, the vector multiplet consists of a gauge field, 2 Weyl fermions and 2 scalars $({\cal A}_\mu, \lambda_{\alpha A}, \bar\lambda_{\dot\alpha A}, \phi,\bar\phi)$ and the hypermultiplet consists of 4 scalars and 2 Weyl fermions $(q_{A I}, \psi_{\alpha I}, \bar\psi_I^{\dot\alpha})$ ($I = 1,2$). 
From Table~\ref{3charges}, the complex scalar $\phi$ (identified as $\phi_3$) is only charged under U(1)$_3$, which can then be identified as U(1)$_R$. \footnote{Note that, in order to turn on U(1)$_R$, we have to temporarily relax the reality condition for $\phi$ and $\bar\phi$.} We note that the charged complex scalar has U(1)$_R$ charge $+2$ and $k_3 = +1$. As a result of the different normalization, the gauge fields are related in the following way
\be
\el{match3}
A^3 = 2  A^{\zt U(1)_R}\ .
\ee
The scalars $q_{A I}$ transform as a doublet of SU(2)$_R$ and they have opposite charges under Cartan subgroup U(1)$_J$ of SU(2)$_R$. We can identify $q_{1 1}$
as $\phi_1$  and $q_{2 1}^\dagger$ as $\phi_2$. From Table~\ref{3charges}, these two scalars have charges $k_1 = +1$ and $k_2 = +1$ respectively. Hence we can fix the coefficients of the linear combination
\be
\el{match2}
 A^{U(1)_J} = \frac 1 2 (A^1 + A^2)\ .
\ee
There is another combination $\frac 1 2 (A^1 - A^2)$, which is not in the
$\CN=2$ R-symmetry group. So the current $\CN=2$ background corresponds to the case of $A^1 = A^2$. Combining these, we can see that the singular limit of background configuration (\ref{diagonalB}) give
\be
\el{2chbk}
A^1 = A^2 = {q-1\over 2}\ ,
\ee
which precisely agrees with (\ref{chemical2}). 

\subsubsection{Localization on resolved branched four-sphere}
\label{subsec:Localization}
The background (\ref{TSbackground})(\ref{solsp}) allows Killing spinor solutions \er{Kspinors}
on the resolved branched sphere. With the corresponding supercharge $\df Q$, we can compute the partition function of $\CN=2$ supersymmetric gauge theories using localization techniques. Since
the procedure is insensitive to the resolving factor $f_\epsilon(\theta)$  and the specific forms of the background
fields ${\bf\tilde V}, {\bf T},{\bf\bar T}, M$, it will be essentially identical to what is presented in \cite{Hama:2012bg}. So we will be as brief as we can and only list the key steps and the final results. Readers interested in the details can consult \cite{Hama:2012bg} (see also \cite{Pestun:2007rz}).

{\bf Saddle point}~~~ First consider the $\CN=2$ vector multiplet, which contains a gauge field ${\cal A}_\mu$, gauginos $\lambda_{\alpha A}, \bar\lambda_{\dot\alpha A}$, two real scalar fields $\phi,\bar\phi$ and an auxiliary field $D_{AB}=D_{BA}$. All of them are Lie algebra valued and satisfy reality conditions. The Lagrangian of supersymmetric Yang-Mills theory on the resolved sphere takes the following form
\begin{eqnarray}
{\cal L}_\text{YM} &=&
\Tr \Big[ 
 \frac12F_{\mu\nu}F^{\mu\nu}
+16F_{\mu\nu}(\bar\phi T^{\mu\nu}+\phi\bar T^{\mu\nu})
+64\bar\phi^2 T_{\mu\nu}T^{\mu\nu}+64\phi^2\bar T_{\mu\nu}\bar T^{\mu\nu}
 -4D_\mu\bar\phi D^\mu\phi\nonumber \\ && +2M\bar\phi\phi 
 -2i\lambda^A\sigma^\mu D_\mu\bar\lambda_A
 -2\lambda^A[\bar\phi,\lambda_A]
 +2\bar\lambda^A[\phi,\bar\lambda_A]
 +4[\phi,\bar\phi]^2
 -\frac12D^{AB}D_{AB}
 \Big]\ .\nn
\label{LYM}
\end{eqnarray}
It is argued in \cite{Hama:2012bg} that, the saddle
point locus on the deformed sphere remains the same as that on a round sphere and it is given by (except at north and south poles),
\begin{equation}
 {\cal A}_\mu =0,\quad
 \phi=\bar\phi=-\frac i2a_0\,,\quad
 D_{AB}= -ia_0w_{AB}\ ,\label{locus}
\end{equation}
where 
\be
w_{AB} \equiv
    \frac{4\xi_A\sigma^{\mu\nu}\xi_B\,(T_{\mu\nu}-S_{\mu\nu})}{\xi^C\xi_C}
 = -\frac{4\bar\xi_A\bar\sigma^{\mu\nu}\bar\xi_B\,(\bar T_{\mu\nu}-\bar S_{\mu\nu})}
         {\bar\xi_C\bar\xi^C}\,.
\ee
Note that the constant matrix $a_0$ needs to be integrated over the Lie algebra
but the integration domain can be reduced to the Cartan subalgebra, contributing an extra factor of Vandermonde determinant. At the north (south) pole, the field strength can take the anti-self-dual (self-dual) form,
leading to instanton (anti-instanton) contribution. The classical contribution to the path integral which follows from evaluating Yang-Mills action (\ref{LYM}) on the locus (\ref{locus}) is given by
\be\label{classicalS}
S = \frac1{g_\text{YM}^2}
 \int \rmd^4x\sqrt g{\cal L}_\text{YM}\Big|_\text{saddle point} =
 \frac{8\pi^2}{g_\text{YM}^2}\,q \ell^2\Tr(a_0^2)\ .
\ee

{\bf One-loop determinant}~~~The value of path integral is invariant under the $\widehat{\df Q}$-exact deformation ${\cal L}\rightarrow {\cal L}+t\widehat{\df Q}{\cal V}^\prime$. \footnote{$\widehat{\df Q} = \df Q + \df Q_B$ with $\df Q_B$ being the BRST operator. ${\cal V}^\prime={\cal V}+{\cal V}_\text{GF}$, where ${\cal V}_\text{GF}$ is the gauge fixing term.} By choosing the bosonic part of $\widehat{\df Q}{\cal V}^\prime$ positive definite and sending $t\rightarrow\infty$, Gaussian approximation becomes exact for the path integral over the fluctuations around the locus.
The Gaussian integral gives the square root of the ratio between the determinant of fermionic kinetic operator $K_\text{fermion}$ and that of the bosonic kinetic operator $K_\text{boson}$,
both of which follow from the quadratic part of the $\widehat{\df Q}$-exact regulator $\widehat{\df Q} {\cal V}^\prime$. The quadratic part of ${\cal V}^\prime$ can be written as
\begin{equation}
{\cal V}^\prime\Big|_\text{quad.} =\left( {\cal V}+{\cal V}_\text{GF}\right)\Big|_\text{quad.} ~=~
 (\widehat{\df Q} X, \Xi)\left( \begin{array}{cc}
 D_{00} & D_{01} \\ D_{10} & D_{11} \end{array}\right)
 \left(\begin{array}{c}
 X \\ \widehat{\df Q}\Xi \end{array}\right)\ ,
\end{equation}
where $D_{ij}$ are differential operators. $X$ and $\Xi$ are bosonic and fermionic fields, respectively.
The fields $X, \Xi$ can be regarded as sections of bundles $E_0, E_1$ on the manifold and therefore $D_{10}$ acts on the complex
\be 
\xG(E_0) \to \xG(E_1)\ .
\ee
The ratio of the determinants can be related to the spectrum of operator $\df H \equiv
\widehat {\df Q}{}^2$ on the kernel and cokernel of the operator $D_{10}$ \cite{Pestun:2007rz}
\begin{equation}
 \frac{\text{det}K_\text{fermion}} {\text{det}K_\text{boson}} ~=~
 \frac{\text{det}_{\text{Coker}D_{10}}\df H}
      {\text{det}_{\text{Ker}D_{10}}\df H}\ .
\end{equation}
The latter can be extracted from the equivariant index of
the transversely elliptic operator $D_{10}$,
\begin{equation}\label{indD10}
 \text{ind}D_{10} ~\equiv~
 \Tr_{\text{Ker}D_{10}}\big(e^{-i\df H t}\big)
-\Tr_{\text{Coker}D_{10}}\big(e^{-i\df H t}\big).
\end{equation}
Here is the explanation how this can actually be done. The determinant of $\df H$ is given by the product of its eigenvalues $\prod_i \xl_i$, while the trace $\Tr e^{-i\df H t}$ can be written as $\sum a^{\xl_i}$ with $a \equiv e^{-i  t}$. To get the determinant, we can take down the exponents
and replace the sum by product. 

According to Atiyah-Bott formula, the index can be evaluated as the sum of contributions from the two fixed points (north and south poles, where $x = \tilde x$)
\begin{equation}\label{indexD10}
 \text{ind}(D_{10})~=~ \sum_{x:\,\text{fixed point}}
 \frac{\Tr_{E_0}\gamma-\Tr_{E_1}\gamma}
      {\text{det}(1-\partial\tilde x/\partial x)}\,.
\end{equation}
Note that generic fields (sections) on the manifold transform under $e^{-i\df H t}$ as
\be e^{-i\df H t}s(x) = \xg_s \,s(\tilde x) .\ee

{\bf Square of fermionic symmetry}~~~As we can see, the index and therefore the one-loop determinant only depend on how the fields and points on the manifold transform under $\df H$. Now we study the action of $\df H=\widehat {\df Q}{}^2$.
The square of fermionic symmetry $\widehat {\df Q}{}^2$ is a linear combination of various
symmetry transformations
\begin{eqnarray}
 \widehat {\df Q}{}^2 &=& i{\cal L}_v + \text{Gauge}(a_0)
 + \text{Lorentz}(L_{ab})
 \nonumber \\ 
 &+& \text{Scale}(w)
 + \text{R}_{\zt U(1)_R}(\Theta)
 + \text{R}_{\zt{SU}(2)_R}(\hat\Theta_{AB})\ ,
\label{hatQ2}
\end{eqnarray}
where the transformation parameters can be read off from the SUSY transformation rule shown in \cite{Hama:2012bg},
\begin{eqnarray}
\label{susyalg1}
 v^\mu & = & 2\bar\xi^A\bar\sigma^\mu \xi_A\ ,\\
 L_{ab} & = & D_{[a}v_{b]}+v^\mu\Omega_{\mu ab}\ ,\\
 w &=&-(i/2) \left(\xi^A\sigma^\mu D_\mu\bar\xi_A+ D_\mu\xi^A\sigma^\mu\bar\xi_A\right)\ ,\\
 \Theta &=& -(i/4) \left(\xi^A\sigma^\mu D_\mu\bar\xi_A - D_\mu\xi^A\sigma^\mu\bar\xi_A\right)\ ,\\
 \hat\Theta_{AB} &= & -i\xi_{(A}\sigma^\mu D_\mu\bar\xi_{B)}+iD_\mu\xi_{(A}\sigma^\mu\bar\xi_{B)}+v^\mu V_{\mu AB}\ .
\end{eqnarray}
The effect of gauge transformation will be discussed later and we will temporarily
take the gauge group to be Abelian. Killing vector $v^\mu\partial_\mu$ as a bilinear of $\bar\xi^A, \xi_A$ can be computed
\begin{equation}
 2\bar\xi^A\bar\sigma^\mu\xi_A\,\partial_\mu ~=~
  \frac{1}{q\ell}\partial_\tau
 +\frac 1\ell\partial_\chi.
\label{kv2}
\end{equation}
Plugging in $\Omega_{nab}$ given in \er{spinconS4} one can further show $L_{ab}=
0$. From the explicit Killing spinor solutions \er{Kspinors}, 
we get $w=\Theta=0$. 
With the use of the main equation \er{ks1}, we can express the $\zt{SU}(2)_R$ parameter $\Theta^A_{\;~B}$ in terms of $S_{ab},\bar S_{ab}$ and $V^A{}_B$. Substituting in \er{bkg3d} and \er{solsp}, we get
\begin{equation}
\hat\Theta^A_{\;~B}= \Big(-\frac1{2q\ell}-\frac1{2\ell}\Big)\cdot
 (\tau^3)^A_{~B}.
\end{equation}
In summary, the action of $\df H$ on the resolved branched space is essentially identical to that on an ellipsoid. This is particularly clear near the poles where $\rho = 0, \pi$. As we can see from the behavior of $f_\xe (\xt)$ (\ref{resolvedf}), functions $F,G,H$ given in \er{metriccomp} return to $f,g,h$ (as a result of $f_\xe \rightarrow q \ell$) and $\widehat \BS_q^4$ turns into an ellipsoid.

For non-Abelian gauge group $G$, $a_0$ is in the Cartan subalgebra and there is an extra factor for the index (\ref{indexD10})
\be
\text{rank}\, G + \sum_{\alpha\in\Delta} e^{ta_0\cdot\alpha}\ .
\ee
Obviously, this factor is independent of the geometry of the manifold.

{\bf Partition function}~~~To summarize, the one-loop determinant for the vector multiplet should be the same as that on an ellipsoid with deformation parameter $b=\sqrt{q}$,
\be
\el{one-loop-vector}
 \text{Det}_\text{vec} = \sqrt{\frac{\text{det}K_\text{fermion}}{\text{det}K_\text{boson}}} =  \prod_{\alpha\in\Delta_+}\frac
 {\Upsilon_q(i\hat a_0\cdot\alpha)
  \Upsilon_q(-i\hat a_0\cdot\alpha)}
 {(\hat a_0\cdot\alpha)^2}\ ,
\ee where $\hat a_0\equiv \ell\sqrt{q} a_0$ and $\Upsilon_q(x)$ is defined to regularize the following infinite products
\be
\Upsilon_q(x) = \prod_{m,n\ge0}
 \big(m q^{1/2} +n q^{-1/2}+Q-x\big)
 \big(m q^{1/2} +n q^{-1/2}+x\big)\ ,\quad
Q \equiv \sqrt{q} + \frac{1}{\sqrt{q}}\ .
\ee

We can also introduce matter to the theory. The components of the ${\cal N}=2$ hypermultiplet matter are localized at the origin \cite{Hama:2012bg}. The one-loop determinant can be computed the same way as before and the final result should be the same as that on an ellipsoid with $b=\sqrt{q}$. For ${\cal N}=2$ hypermultiplet in representation $R$ we have
\be
\el{one-loop-matter}
 \text{Det}_\text{hyp} =
 \prod_{\rho\in R} \Upsilon_q(i\hat a_0\cdot\rho+\tfrac Q2)^{-1}\ .
\ee

Let us now consider the contribution from the instantons localized at two poles. In the neighborhood of the north pole $x_0 = \ell$ ($\rho = 0$), we can choose the Cartesian coordinates $x_{1,\dots, 4}$ and the metric becomes
flat $g_{\mu\nu} \simeq \eta_{\mu\nu}$ after we drop terms vanishing as $\CO(x^2)$ or
faster. Up to this order, the only nonvanishing background field is the tensor field ${\bf T}$ and it reduces to the value at the pole
\ba
{\bf T} & \simeq & \lsb \frac14\Big(\frac1 F-\frac1 G\Big)\tau^1_\eta
 +\frac H {4 F G}\tau^2_\eta \rsb\at{\rho = 0} 
 =  \frac14\Big(\frac1 f-\frac1 {g}\Big)\tau^1_\eta
 +\frac {h} {4 f g}\tau^2_\eta \at{\rho = 0,\; \tilde \ell = q\ell}\ .
\ea
The second equality follows from the explicit forms of $F,G,H$ and $f,g,h$ (see \er{metriccomp} and \er{ellipfgh} in Appendix~\ref{App:Resolved}). One can show that locally the background fields take the same form
of the Omega background with $\xe_1 = \iv \ell, \xe_2 = {(q \ell)}^{-1}$
\begin{eqnarray}
&&
 {\bf T}^{\Omega}\equiv\frac12 T^{\Omega}_{\mu\nu}\rmd x^\mu\rmd x^\nu =
 \frac1{16}\Big(\frac1{q\ell}-\frac1\ell\Big)
 \big(\rmd x_1\rmd x_2-\rmd x_3\rmd x_4\big)\ ,
 \nonumber \\
&&
{\bf V}^{\Omega}={\bf \bar  T}{}^{\Omega}=M^{\Omega}=0\ .
\label{Omega}
\end{eqnarray}
Therefore the instanton contribution is essentially given by the Nekrasov's instanton partition function $Z_{\zt{inst}} (\xe_1, \xe_2, a_0, \tau)$, where $\tau = \frac \xt {2\pi} + \frac {4\pi}{g_{\zt{YM}}^2} i$.
Similarly, we get instanton contribution from the south pole $Z_{\zt{inst}} (\xe_1, \xe_2, a_0, \bar\tau)$.

Putting all the pieces together, the partition function on the resolved sphere is
\begin{eqnarray}
 Z =
 \int \prod_i \rmd (\hat a_0)_i\,
 e^{-\frac{8\pi^2}{g_\text{YM}^2}\Tr(\hat a_0^2)}
 \frac{\prod_{\alpha\in\Delta_+}
 \Upsilon_q( i\hat a_0\cdot\alpha)
 \Upsilon_q(-i\hat a_0\cdot\alpha)}
 {\prod_{\cal I}\prod_{\rho\in R_{\cal I}}\Upsilon_q(i\hat a_0\cdot\rho+\tfrac Q2)}  |Z_\text{inst}|^2\ ,
\label{Ztot}
\end{eqnarray} where ${\cal I}$ denotes different types of hypermultiplet matter.
Note that the partition function is independent of the resolving function $f_\epsilon$
and therefore we can take the limit $\xe \to 0$ and obtain the partition function on the branched sphere $\mathbb{S}_q^4$
\be
Z_q = Z_{\epsilon\rightarrow 0}= Z\ .
\ee

\subsubsection{Other supersymmetric backgrounds}
\label{otherSUSYbks}
It is not clear to us how localization can be performed in a generic supersymmetric background \eqref{chemical4}, which also allows Killing spinors. We will instead assume that localization can be done and present a somewhat ad hoc method to compute the one-loop determinant for $\CN=4$ SYM. We propose that one-loop determinant can be obtained by simply shifting the unmatched fermionic and bosonic eigenvalues according to the change of U(1) background gauge fields. It is natural to expect the classical contribution to remain the same. The instanton contribution is more subtle but fortunately it can be neglected in the large $N$ limit. 

Our strategy is to consider the contribution from each $\CN=1$ multiplet. The $\CN=2$ hypermultiplet consists of two $\CN=1$ chiral multiplets while the $\CN=2$ vector multiplet consists of one chiral and one vector multiplets from the $\CN=1$ point of view. Note that the R-charge $r$ of each ${\cal N}=1$ multiplet \footnote{For simplicity, we will continue to use the term R-charge even though it is essentially a linear combination of the R-charge and the flavor charges.} is determined by 
\be\el{n1rcharge} \lb \frac {q-1} 2\rb r = k_i A^i\ .\ee 
It means that we consider a single effective background gauge field and the associated charges of the surviving Killing spinors is $r = \pm 1$. 

Let us start from the known index of a ${\cal N}=2$ hypermultiplet \cite{Hama:2012bg} on the background \er{2chbk}, in which each ${\cal N}=1$ multiplet has R-charge $r=1$ and contributes to the index ($q_1 = e^{\frac {i\,t} \ell}, q_2 = e^{\frac {i\,t} {q \ell}}$),
\begin{eqnarray}
\el{indhyper}
 \text{ind}(D_{10}^\text{r=1})=\frac 1 2 \text{ind}(D_{10}^\text{hyp}) &=&
  \left[\frac{q_1{}^{1/2}q_2{}^{1/2}}{(1-q_1)(1-q_2)}\right]_+
 +\left[\frac{q_1{}^{1/2}q_2{}^{1/2}}{(1-q_1)(1-q_2)}\right]_-
 \nonumber \\ &=&
 \sum_{m,n\ge0}\Big(q_1{}^{m+\frac12}q_2{}^{n+\frac12}
                +q_1{}^{-m-\frac12}q_2{}^{-n-\frac12}\Big)\ ,
\end{eqnarray}
where we use $+$ and $-$ to denote the contributions from the north and south
poles respectively. \footnote{The two terms look identical but they should
be treated in different manners. For example $1/(1-q_1)$ in the first term should
be expanded in power series of $q_1$ while in the second term it should be expanded
in powers of $\iv {q_1}$.} As discussed in sec~\ref{subsec:Localization}, each term in the sum is an eigenvalue of $e^{-i \df H t}$ and the exponent can be regarded as the unpaired eigenvalues of kinetic operators. Under a change of the background fields, the unpaired eigenvalues are shifted. Now we argue that the only change to the eigenvalue is from the covariant derivative $D_\mu$. For simplicity we will continue to consider the index instead of the product of eigenvalues.

First of all, we rewrite \er{indhyper} as
\be
\text{ind}_+(D_{10}^\text{r=1}) = \frac{q_1{}^{1/2}q_2{}^{1/2}}{(1-q_1)(1-q_2)} = \frac{q_1{}^{1/2}q_2{}^{1/2}-q_1{}^{3/2}q_2{}^{1/2}}{(1-q_1)^2(1-q_2) }\ ,
\ee
where $\text{ind}_+$ denotes the contribution from the north pole. \footnote{For simplicity we only consider one pole and the contribution from south pole is formally identical.} The index is recast into this form since we assume the general one-loop
determinant is given by triple Gamma functions. The first term in the numerator $\sqrt{q_1q_2}$ corresponds to the contribution from the unmatched eigenmodes of the complex scalar while the second term is from the spinor. We notice that the background field in the covariant derivative contributes a phase to $e^{-i \df H t}$
\[\sqrt {\frac {q_2} {q_1}} = e^{i\lb \frac 1 {2q \ell} - \frac 1 {2\ell} \rb
t}\,.\]
As a result, a change from background \er{2chbk} to background \eqref{chemical1} reduces the
R-charge of the ${\cal N}=1$ chiral multiplet by one and leads to the following change,
\be
\sqrt {q_1q_2} \to q_1,\quad q_1{}^{3/2}q_2{}^{1/2} \to q_1 q_2\ .
\ee
The index then becomes
\be
\label{N1r0}
 \text{ind}_+(D_{10}^\text{r=0})=\frac{q_1-q_1q_2}{(1-q_1)^2(1-q_2)} = \frac{q_1}{(1-q_1)^2 }\ ,
\ee
which is equal to half of the index of a $\CN=2$ hypermultiplet on a round sphere. 
Generally, we conjecture that the index of a chiral superfield with R-charge $r$ is given by
\be
\el{chiralmultiplet}
 \text{ind}_+(D_{10}^\text{r})=\frac{q_1 \lb \frac {q_2} {q_1}\rb^{\frac r 2}-q_1^2\lb \frac {q_2} {q_1}\rb^{1-\frac r 2}}{(1-q_1)^2(1-q_2) }\ .
\ee

Now we turn to $\CN=2$ vector multiplet. The known index \cite{Hama:2012bg} on the background \er{2chbk} can be decomposed as \footnote{Here we drop the contribution from constant modes (equal to $+2$) to the total index since it remains the same in all different cases. We will recover it for the one-loop determinant in the end.}
\be
\text{ind}_+(D_{10}^\text{vec})=\frac{-1-q_1q_2}{(1-q_1)(1-q_2)} = \frac{q_1^2 q_2-1}{(1-q_1)^2 (1-q_2)}+\frac{q_1-q_1q_2}{(1-q_1)^2(1-q_2)}\ .
\ee
The second term is the contribution from the ${\cal N}=1$ chiral multiplet of R-charge $r=0$.
The first term is the contribution from the $\CN=1$ vector multiplet and it remains the same in different backgrounds. 

{\bf A single U(1)}~~~In the case of background \eqref{chemical1}, the contribution from the chiral multiplet of ${\cal N}=2$ vector multiplet follows from \er{chiralmultiplet} with $r=2$
\be
 \text{ind}_+(D_{10}^\text{r=2})=\frac{q_2-q_1^2}{(1-q_1)^2 (1-q_2)}\ ,
\ee
and the full index of the ${\cal N}=2$ vector multiplet becomes,
\be
 \text{ind}_+(D_{10}^\text{vec})=\frac{-1-q_1^2}{(1-q_1)^2}\ ,
\ee
which is the same as the index on a round sphere. The index of the ${\cal N}=2$ hypermultiplet is given by twice of \er{N1r0} and we get that the total index of ${\cal N}=4$ multiplet is $-1$, which is exactly the same as the index on $\BS^4$.
Therefore, the one-loop determinant of ${\cal N}=4$ SYM on the branched sphere with a single $U(1)$ background field is identical to that on a round sphere.

{\bf Three U(1)'s}~~~In the generic background \eqref{chemical4}, the R-charges of the three chiral multiplets are $+\frac{2 a }{3}$, $+\frac{2 b }{3}$ and $2-\frac{2 a +2b}{3}$ (following from \er{n1rcharge}). The total index reads,
\ba
 \text{ind}_+(D_{10}^\text{vec+hyp})& =& \frac{q_1^2 q_2-1}{(1-q_1)^2 (1-q_2)} + \frac{ q_1{}^{1-\frac a 3} q_2{}^{\frac a 3} -q_1{}^{1+\frac a 3} q_2{}^{1-\frac {a} 3}}{(1-q_1)^2 (1-q_2)}\nn
& & +\frac{ q_1{}^{1-\frac b 3} q_2{}^{\frac b 3} - q_1{}^{1+\frac b 3} q_2{}^{1-\frac b 3}}{(1-q_1)^2 (1-q_2)} +\frac{ q_1{}^{\frac {a+b} 3} q_2{}^{1-\frac {a+b} 3} -q_1{}^{2-\frac {a+b} 3} q_2{}^{\frac {a+b} 3}}{(1-q_1)^2 (1-q_2)}\ .
\ea
Expanding the denominator in power series and then replacing the sum by a
product, we get the one-loop determinant,
\be
\zt{Det}_\text{vec+hyp} = \prod_{\alpha\in\Delta_+} \frac 1 {(\hat a_0\cdot\alpha)^2} \frac{G(2,1)\, G(\frac {3-a} 3,\frac{a}{3})\, G(\frac {a+b}{3}, \frac{3-a-b}{3})\, G(\frac {3-b} 3, \frac{b}{3})}{G(0,0)\, G(\frac{3+a}{3}, \frac{3-a}{3})\, G(\frac{6-a-b}{3}, \frac{a+b}{3})\, G(\frac{3+b}{3}, \frac{3-b}{3})}\ ,\el{one-loop_case4}
\ee where the function $G(x, y)$ is defined by the product of two triple Gamma functions with $\vec \xo_3 = (q^{1/2},q^{1/2},q^{-1/2})$, \footnote{See Appendix \ref{MGamma} for the definitions of multiple Gamma functions.}
\be
G(x, y) = \Gamma _3\left(x q^{1/2}+ y q^{-1/2}+i\hat a_0\cdot\alpha,\vec \xo_3\right)\Gamma _3\left(x q^{1/2}+y q^{-1/2}-i\hat a_0\cdot\alpha,\vec \xo_3\right)\ .
\ee
In the particular case with $a=b=1$, (\ref{one-loop_case4}) reduces to
\ba
\el{one-loop_case3}
\zt{Det}_\text{vec+hyp} & = & \prod_{\alpha\in\Delta_+} \frac 1 {(\hat a_0\cdot\alpha)^2}\frac {G(2,1)}{G(0,0)}\times \lsb \frac {G(\frac 23, \frac 13)}{G(\frac 43, \frac 23)}\rsb^3\ .
\ea
We will see that both (\ref{one-loop_case4}) and (\ref{one-loop_case3}) have simple behaviors, to the leading order in the large eigenvalue expansion. 
\subsubsection{Partition function in the large $N$ limit}
In Section \ref{subsec:Localization}, we have shown that the path integral of ${\cal N}=2$ gauge theory on branched sphere $\mathbb{S}_q^4$ with two U(1) background fields (\ref{2chbk}) can be localized in the Coulomb branch to a finite-dimensional matrix integral. We are particularly interested in the special case of ${\cal N}=4$ theory with hypermultiplet in the adjoint representation of gauge group SU($N$). Our goal in this section is to study the resulting matrix model in the large $N$ limit. By $\hat a_0\cdot\rho=\hat a_0\cdot\alpha$, the matrix integral (\ref{Ztot}) of ${\cal N}=4$ theory can be written as
\begin{eqnarray}
 Z =
 \int \prod_i \rmd (\hat a_0)_i
 e^{-\frac{8\pi^2N}{\lambda}\Tr(\hat a_0^2)}
 \prod_{\alpha\in\Delta_+}
\frac {\Upsilon_q( i\hat a_0\cdot\alpha)
 \Upsilon_q(-i\hat a_0\cdot\alpha)}
 {\Upsilon_q(i\hat a_0\cdot\alpha+\tfrac Q2)
 \Upsilon_q(-i\hat a_0\cdot\alpha+\tfrac Q2)}  |Z_\text{inst}|^2\ ,
\label{Ztot1}
\end{eqnarray} where $\lambda=g_\text{YM}^2N$ is the 't Hooft coupling and the instanton contributions $ |Z_\text{inst}|^2$ become negligible at large $N$ due to exponential suppression~\cite{Russo:2012ay}. From now on we will set $Z_\text{inst}=1$.

In the planar limit, the matrix integral (\ref{Ztot1}) is governed by the saddle point. In terms of the eigenvalue density
\be
\rho(x) = {1\over N}\sum_i \delta (x - (\hat a_0)_i)\ ,
\ee the saddle point equations are equivalent to a singular integral equation
\be
 \strokedint_{-\mu }^{\mu } dy\,\rho (y) K(x-y) = \frac{8\pi ^2}{\lambda }\,x\ .\label{intsadd}
\ee
The function $K(x)$ here is defined as
\be
K(x) = \frac12\partial_x \log\left(\frac {\Upsilon_q(ix)
 \Upsilon_q(-ix)}
 {\Upsilon_q(ix+\tfrac Q2)
 \Upsilon_q(-ix+\tfrac Q2)}\right)\ .
\ee Recall that $\Upsilon_q(x)$ can be decomposed as Barnes double gamma functions
\begin{align}
\Upsilon_q(x)& =\prod_{m,n\ge0}\big(m q^{1/2} +n q^{-1/2}+x\big) \big(m q^{1/2} +n q^{-1/2}+Q-x\big) \nn
 & = \frac{1}{\xG_2[x, (q^{1/2}, q^{-1/2})]~\xG_2[Q-x,(q^{1/2}, q^{-1/2})]}\ .
\end{align}
At large $|x|$, Barnes double gamma function can be expanded as \footnote{The large $x$ expansion of $\log \xG_n(x,\vec \xo)$ is given in Appendix \ref{MGamma}.}
\ba
\log \xG_2[x,(a,b)] &=& -\frac{1}{2 a b}x^2 \log x + \frac{3}{4 a b} x^2 +\frac{1}{2}  \left(\frac{1}{a}+\frac{1}{b}\right) (x\log x-x)\nn
& &-\left(\frac{1}{12} \left(\frac{a}{b}+\frac{b}{a}\right)+\frac{1}{4}\right)\log x +\cdots\ .\el{G2lartexexp}
\ea
Then at large $x$, $K(x)$ becomes 
\be
K(x)= {(1+q)^2\over 4q}{1\over x} + {(q^2-1)^2\over 96 q^2}{1\over x^3} + {\cal O}(x^{-4})\ .\label{xexpandK}
\ee
When $q\rightarrow1$, all the higher terms vanish, $K(x)$ becomes ${1\over x}$ and the saddle point equation (\ref{intsadd}) returns to that of ${\cal N}=4$ SYM on round sphere $\mathbb{S}^4$
\be
 \strokedint_{-\mu }^{\mu } dy\,\rho (y) {1\over x-y} = \frac{8\pi ^2}{\lambda }\,x\ .\label{intsaddN=4}
\ee
To leading order in the large $x$ expansion (\ref{xexpandK}), the $q$-dependence of $K(x)$ is simply factorized
\be
K(x)\approx {Q^2\over 4} {1\over x}\ , \quad Q = \sqrt{q}+\frac{1}{\sqrt{q}}\ .
\ee
Notice that $ \int dy\,\rho (y)$ is always order one and therefore the large $x$ expansion is essentially the large $\xl$ expansion by requiring consistent scalings of $x$ and $\lambda$ in the saddle point equation.

From now on we take this leading order approximation and then the saddle point equation (\ref{intsadd}) becomes that of ${\cal N}=4$ SYM on $\mathbb{S}^4$ with a rescaled 't Hooft coupling
\be
 \strokedint_{-\mu }^{\mu } dy\,\rho (y) {1\over x-y} = \frac{8\pi ^2}{\widetilde\lambda }\,x\ ,\quad \widetilde\lambda={Q^2\over 4}\lambda\ .\label{intsaddq}
\ee This saddle point equation (\ref{intsaddq}) is solved by Wigner's semicircle
\be
\el{saddledensity}
\rho(x) = {8\pi\over \widetilde\lambda} \sqrt{\mu^2-x^2}\ ,
\ee where the width $\mu$ is determined by the normalization condition
\be
1={4\pi^2\mu^2\over \widetilde\lambda}\ ,\quad\mu=\frac{ \sqrt{\widetilde\lambda}}{2\pi}=\frac{ \sqrt{\lambda}}{4\pi}Q\ .
\ee
With this solution, the large $N$ free energy on $\mathbb{S}^4_q$ can be computed by
\ba
F_q&=&-\log Z_q\nn
&=& \frac {8 \pi^2 N^2} \xl \int_{-\mu }^{\mu } \rho(x) x^2 \rmd x - \frac {N^2}2\frac {Q^2} 4 \int_{-\mu }^{\mu } \rho(x)\strokedint_{-\mu }^{\mu } \rho(y)\log (x-y)^2 \rmd x \rmd y\,.~~\el{largeNactionq}
\ea
The first term of (\ref{largeNactionq}) is evaluated to be
\be
\frac {8 \pi^2 N^2} \xl \int_{-\mu }^{\mu } \rho(x) x^2 \rmd x = \frac12 N^2{\widetilde\lambda\over\lambda}\ .
\ee
Using the identity
\be
\strokedint_{-\mu}^\mu \sqrt{\mu^2-y^2}\,\log|x-y| \rmd y = {\pi\over 2}\left(x^2 - {\mu^2\over 2} + \mu^2\log{\mu\over 2}\right)
\ee to simplify the second term of (\ref{largeNactionq}), the final relevant log term of free energy can be obtained
\be
F_q = -\frac12 N^2{\widetilde\lambda\over\lambda} \log\widetilde\lambda=-\frac12 N^2{Q^2\over 4}\log\widetilde\lambda\ .\label{Flog}
\ee One can check that, at $q=1$, $\widetilde\lambda=\lambda$ and (\ref{Flog}) is exactly the result of ${\cal N}=4$ SYM on round sphere. In the strong 't Hooft coupling limit, the $q$-dependence inside the log in (\ref{Flog}) is negligible and therefore the $q$-dependence of free energy $F_q$ simply factorizes
\be
F_q = {Q^2\over 4} F_1 = \frac14\left(\sqrt{q}+{1\over \sqrt{q}}\right)^2 F_1\ .
\ee The SRE is then obtained as
\be
{S_q\over S_1} = {3q+1\over 4q}\ .\label{f2_localization}
\ee
Both free energy and SRE precisely agree with the results of free field computation, which implies that both of them are protected. Indeed the coefficient of the log in the $q\to1$ limit is associated with the Weyl anomaly and independent of the coupling. Our exact result (\ref{Flog}) suggests that the universal part of the free energy on $q$-branched sphere $\mathbb{S}^4_q$ is also independent of the coupling constant.

Now let us turn to the partition functions in the other backgrounds, with a single U(1) field (\ref{chemical1}) and three U(1) fields (\ref{chemical4}) turned on, respectively.  In either case, classical part should be the same as (\ref{classicalS}) since it does not depend on the R-charge coupling. The one-loop determinant in the first case remains the same as on a round sphere, as discussed in Section \ref{otherSUSYbks}. Neglecting the instanton contribution in the large $N$ limit, the saddle point equation takes the
form of ${\cal N}=4$ SYM on $\mathbb{S}^4$ (\ref{intsaddN=4}).  
Therefore one can easily see that both the free energy and SRE are $q$-independent
\be
F_q = F_1\ ,\quad S_q = S_1\ ,\label{f1_localization}
\ee which agrees with the free field computation (\ref{freeSq1}).

Now we move on to the background \eqref{chemical4} with three U(1) chemical potentials turned on. The one-loop determinant in this case is given by \er{one-loop_case4}. To the leading order in large $x$ expansion, the kernel $K(x)$ in equation (\ref{intsadd}) now becomes \footnote{The large $x$ expansion of triple Gamma function can be found in \er{G3lartexexp}.}
\be
K(x)\approx \frac{[a (q-1)-3 q] [b (q-1)-3 q] [a (q-1)+b (q-1)+3]}{27 q^2}{1\over x}\ ,
\ee
and consequently the free energy has the following scaling behavior
\be
F_q = \frac{[a (q-1)-3 q] [b (q-1)-3 q] [a (q-1)+b (q-1)+3]}{27 q^2} F_1\ .\label{f3_localization}
\ee
It is not difficult to show that the SRE scales exactly the same as \er{Sq3charges}.
In the special case of three equal U(1) fields ($a=b=1$), $q$ scaling of the free energy now becomes
\be
F_q = \frac{(2 q+1)^3}{27 q^2} F_1\ ,
\ee
and the SRE scales like
\be
{S_q\over S_1}  = \frac{19 q^2+7q+1}{27 q^2}\ .
\ee
As we can see, in every case the strong coupling results (\ref{f1_localization})(\ref{f2_localization})(\ref{f3_localization}) precisely agree with the free field results (\ref{freeSq1})(\ref{Sq2charges})(\ref{Sq3charges}).

\section{Five-dimensional R-charged Topological Black Hole}\label{sec:TBH}

Now we search for gravity duals for the four-dimensional superconformal field theories on $\mathbb{S}^4_q$. As discussed before, the rigid supersymmetry on $\mathbb{S}^4_q$ requires additional background fields, which couple to the conserved R-currents. For simplicity, we want to restrict ourselves to Abelian R-currents. The R-symmetry group of ${\cal N}=4$ super Yang-Mills is $\zt{SO}(6)$ and its maximal Abelian subgroup is the Cartan $\zt U(1)\times \zt U(1)\times \zt U(1)$. Adding R-symmetry backgrounds ( physically interpreted as chemical potentials ) in field theory corresponds to adding R-charges on the gravity side. 
Due to the conical singularity on $\mathbb{S}^4_q$, it is easier to search for gravity duals for field theories on the conformally equivalent space
$\BS_q^1\times \BH^3$.  In this section, we focus on the candidates for the gravity duals, which are the charged AdS topological black hole solutions in five-dimensional ${\cal N}=2$ STU gauged supergravity theory.

\subsection{Five-dimensional $\CN=2$ gauged supergravity}
Five-dimensional $\CN=2$ supergravity theories can be realized as eleven-dimensional supergravity compactified on Calabi-Yau three-folds~\cite{Papadopoulos:1995da,Cadavid:1995bk}. The massless spectrum
of the compactified theory contains $n_V=h_{(1,1)}-1$ vector multiplet and $n_H = h_{(2,1)}+1$ hypermultiplet, where $h_{(1,1)}$ and $h_{(2,1)}$ are Hodge numbers of the Calabi-Yau manifold. For our purpose the hypermultiplets are switched off. The field contents of the supergravity multiplet are the f\"unfbein $e_\mu^a$, two gravitini
$\psi_\mu^A$ and a graviphoton $\widetilde A_\mu$. Each vector multiplet contains a
vector $A_\mu$, two spinors $\xl^{ A }$ and one real scalar $\phi$. The fermions
in each multiplet transform as doublet (label by the superscript $A$) under the SU(2)$_R$ R-symmetry group
while all the other fields are neutral. Anti-de Sitter solutions can be obtained
by gauging the U(1) subgroup of SU(2)$_R$. This is done by introducing 
coupling to a linear combination of the $n_V+1$ (including the graviphoton) Abelian gauge fields 
\be
V_I A_\mu^I\ ,\quad I = 1\dots n_V+1
\ee with coupling
constant $g={1\over L}$. The bosonic part of the gauged supergravity Lagrangian is given by
\be
{\mathcal{L}\over \sqrt{-g}} = -\frac{1}{2} R + {V\over L^2} - \frac{1}{4} G_{IJ}F_{\mu\nu}{}^I F^{\mu\nu J}-\frac{1}{2} g_{ij}\partial_{\mu}\phi^i\partial^\mu\phi^j+\frac{1}{48\sqrt{-g}}\epsilon^{\mu\nu\rho\sigma\lambda} C_{IJK}F_{\mu\nu}^IF_{\rho\sigma}^JA_\lambda^K \ ,\label{action}
\ee where $i=1\dots n_V$ and $V$ is the scalar potential given by \footnote{The scalar potential is necessary because of supersymmetry.}
\be\label{potential}
V = V_I V_J\Big( 6X^I X^J - {9 \over 2} g^{ij} \partial_{i}X^I\partial_{j}X^J\Big)\ .
\ee
The real scalar fields $X^I$ have to satisfy the constraint,
\be
{\cal V} = {1 \over 6} C_{IJK} X^I X^J X^K =1.
\ee
The homogeneous cubic polynomial $\CV$ specifies a hypersurface embedded in the $n_V+1$-dimensional space parameterized by $X^I$ and this hypersurface is the target space $\CM$ with $\xvp^i$ as coordinates \footnote{In other words, $X^I$ are known functions of $\phi^i$ and these functions themselves are arbitrary as the Lagrangian is invariant under redefinition
of $\phi^i$.}. This manifold is known as ``very special" manifold. Other quantities in \er{action}, $G_{IJ}$ and $g_{ij}$ can be expressed in terms of $\CV$,
\begin{equation}
G_{IJ} = -{\frac{1}{2}}{\frac{\partial}{\partial X^I}} {\frac{\partial}{%
\partial X^J}}(\ln {\cal V})|_{{\cal V} =1}\ ,\qquad g_{ij} = G_{IJ}
\partial_{i}X^I\partial_{j}X^J|_{{\cal V} =1}\ ,
\label{metric}
\end{equation}
where $\partial_i \equiv {\frac{ \partial }{\partial\phi^i}}$. The matrix
$g^{ij}$ in \er{potential} is the inverse of $g_{ij}$, the latter of which is the metric on $\CM$. 
The BPS solution in the gauged supergravity theory was found in \cite{Behrndt:1998ns}. We leave the general BPS solution and the Killing spinor analysis in Appendix \ref{App:TBH}. In what follows, we will pay our attention to a special case of the gauged supergravity theory, called STU model.
\subsection{STU black hole}
The STU model is a special case of the $\CN=2$ gauged supergravity and it is given by
\be
{\cal V} = X^1X^2X^3 = 1\ .
\ee
Then we get $G_{IJ}$ from (\ref{metric})
\be
G_{IJ}={1\over 2} \left(
\begin{array}{ccc}
 (X^1)^{-2} &   &   \\
   &  (X^2)^{-2} &  \\
  &  &  (X^3)^{-2} \\
\end{array}
\right)\ ,
\ee and with $V_I = \frac13$ we get the potential
\be
V = 2\left ( {1\over X^1}+{1\over X^2}+{1\over X^3}\right)\ .
\ee
The three-charge non-extremal black hole solution is described by the metric
\ba
\rmd s^2 = -{\cal H}^{-4/3} f (r) \rmd t^2 + {\cal H}^{2/3}\left({1\over f(r)} \rmd r^2 + r^2 \rmd\Sigma_{3,k}\right)\ ,\nn
f(r) = k-{m\over r^2} +{r^2\over L^2}{\cal H}^2\ , \quad {\cal H}^2 = H_1H_2H_3\ , \quad H_i = 1+{Q_i\over r^2}\ ,\label{STUbackground}
\ea as well as the scalars and the gauge fields
\be
\el{gaugefield}
X^i = {{\cal H}^{2/3}\over H_i}\ ,\quad A^i = \lsb\sqrt{k+ {m\over Q_i}}\left({1\over H_i}-1\right)-\hat\mu_i\rsb\rmd t\ .
\ee The parameter $k$ specifies the spatial curvature of $\rmd\Sigma_{3,k}$. For flat space $\BR^3$ and three-sphere $\BS^3$, $k$ takes the values of $0$ and $+1$ respectively. For hyperbolic space $\BH^3$, $k=-1$. This particular solution in the STU model is found by Behrnd, Cvetic and Sabran~\cite{Behrndt:1998jd}. This solution with three $\zt U(1)$ charges can also be obtained by $\mathbb{S}^5$-reduction of the ten-dimensional gravity solution coming from spinning D3 branes~\cite{Cvetic:1999xp,Chamblin:1999tk,Cvetic:1999ne}. \footnote{The number of independent angular momenta is exactly the rank of the isometry group $\zt{SO}(6)$ of the six-dimensional space transverse to the branes.
} 

We are particularly interested in the extremal limit $m=0$ and $k=-1$
(boundary being $\BS^1\times \BH^3$). This is a topological BPS black hole as it is a special case of \er{BPSTBH}. Define the rescaled charges $Q_i$ as
\be \kappa_i:={Q_i\over r_h^2}\ ,\ee
where $r_h$ is the largest root of the equation 
\be
f(r_h)=0\ .
\ee Then $\kappa_i$ satisfy the relation\be
(1+\kappa_1)(1+\kappa_2)(1+\kappa_3){r_h^2\over L^2} = 1\ ,
\ee which shows the black hole horizon is determined by the rescaled charges. 
The Hawking temperature of the STU metric (\ref{STUbackground}) can be expressed as
\be
T={1-\kappa_1\kappa_2-\kappa_1\kappa_3-\kappa_2\kappa_3-2\kappa_1\kappa_2\kappa_3\over (1+\kappa_1)(1+\kappa_2)(1+\kappa_3)}T_0\ ,\quad T_0={1\over 2\pi L}\ .
\ee
The Bekenstein-Hawking entropy is given by the outer horizon area
\be
S_{\text{BH}} = {A\over 4 G_5} = {V_3 L^3 \over 4 G_5}{1\over (1+\kappa_1)(1+\kappa_2)(1+\kappa_3)}\ ,
\ee where $V_3$ is the volume of unit hyperbolic space. The three total charges are computed by Gauss law\footnote{It can also be computed by ${1\over 16\pi G_5}\int j^0$, where $j^\mu$ is the conjugate momentum $j^\mu = -\sqrt{g}F^{r\mu}$ for the canonical Maxwell action.}
\be
\widehat Q_i ={V_3\over 8\pi G_5}i Q_i:=V_3~\rho_i\ ,
\ee Here we have taken into account the scalar profile. Using charge-horizon relation, $\widehat Q_i$ can be further expressed as
\be
\widehat Q_i =   {V_3L^2\over 8\pi G_5}{i \kappa_i\over (1+\kappa_1)(1+\kappa_2)(1+\kappa_3)}\ .
\ee The chemical potentials conjugate to the charge densities $\rho_i$ are determined by requiring the gauge potentials vanishing at the horizon $A^i\,|_{r=r_h} = 0$ \footnote{In order to compare with the chemical potential in field theory, one has to take into account the Wick rotation, because so far we proceed in Lorentz signature for black hole.}
\be
\el{chemicalbulk}
\hat\mu_i = A_t^i\,|_{r\to\infty} = {i\over \kappa_i^{-1}+1}\ .
\ee
We have expressed $T, S_{\text{BH}}, \widehat Q_i, \hat\mu_i$ in terms of $\kappa_1,\kappa_2,\kappa_3$ with constant coefficients. It strongly implies that all physical quantities we might compute from this system will solely depend on the rescaled charges. From now on we only use the rescaled charges $\kappa_i$ as variables.

\section{TBH$_{5}$/qSCFT$_4$ Correspondence}
\label{sec:TBH_qSCFT}
In this section we show that the gravity dual of ${\cal N}=4$ super Yang-Mills on branched sphere $\mathbb{S}^4_q$ is the charged topological STU black hole. This correspondence is proposed based on the fact that the R-symmetry background fields on $\mathbb{S}^4_q$, which are necessary to compensate the conical singularity, precisely correspond to the R-charges of the dual black holes. The matching between the $\zt U(1)^3$ bulk gauge fields and the boundary fields is given by
\be
\el{match1}
gA^i_{\text{bulk}}(r\to \infty) = A^{i}_{\mathbb{S}^4_q}\ , \quad i=1,2,3\ .
\ee 
In what follows we shall test the TBH$_5$/qSCFT$_4$ correspondence by comparing supersymmetric R\'enyi entropy and free energy.

We now compute the SRE holographically from the charged topological ($k=-1$) STU black hole specified by (\ref{STUbackground}) (\ref{gaugefield}). As we will see, in
every case, both the SRE and the free energy agree with the localization results as well as the heat kernel computation in the free field limit. Substituting the value of $\kappa_i$ into (\ref{chemicalbulk}), one can see that the TBH chemical potentials and the field theory chemical potentials ( given by \eqref{chemical1},
\eqref{chemical2}, \eqref{chemical3} respectively ) satisfy (\ref{match1}). \footnote{Note that the equality between one forms (\ref{match1}) has included the Wick rotation.}

{\bf A Single charge}~~~
We first consider the STU topological black hole with only one charge,
\be
\el{kappa1}
\kappa_3=\kappa\ ,\quad \kappa_1=\kappa_2 = 0\ .
\ee As discussed before, the system now only depends on a single variable $\kappa$. Since the SRE involves a branching parameter $q$, it will be enough if we figure out the relation between $\kappa$ and $q$. This is obtained by requiring that the Bekenstein-Hawking temperature matches to the geometric period of  the boundary $\mathbb{S}^1$
\be
T = T_0/q\ ,
\ee which gives
\be
\kappa = {q-1}\ .
\ee
Expressing all quantities in terms of the branching parameter, it is convenient to compute SRE using the derived formula in \cite{Huang:2014gca},
\be
\label{REthermal}
S_q = {-q\over q-1}\int_q^1\left( {S_{\text{BH}}(n)\over n^2} - {\widehat Q(n)\hat\mu^{\prime}(n)\over T_0}\right)\rmd n\ .
\ee Evaluating the formula above we get
\be
{S_q\over S_1} =1\ ,\quad S_1= {V_3L^3\over 4G_5}\ . 
\ee
The $q$-independence of SRE implies the $q$-independence of free energy 
\be
I_q :=-\log Z(T,\mu_i) = I_1\ .
\ee
It can be easily checked that $I_1 = -S_1$, which remains valid in all other
cases. 

{\bf Two equal charges}~~~
Now we consider the STU topological black hole with two equal charges,
\be
\el{kappa2}
\kappa_1=\kappa_2=\kappa\ ,\quad \kappa_3 = 0\ .
\ee
The $\kappa-q$ relation is obtained by requiring $T = T_0/q$,
which gives
\be
\kappa = {q-1\over q+1}\ .
\ee
The formula \er{REthermal} can be generalized straightforwardly ($i = 1,2,3$)
\be
S_q = {-q\over q-1}\int_q^1\left( {S_\text{BH}(n)\over n^2} - {\widehat Q_i(n)\hat\mu^{i\prime}(n)\over T_0}\right)\rmd n\ ,
\ee and the SRE is given by
\be
{S_q\over S_1} ={\frac{3q+1}{4 q}}\ . 
\ee
The $q$-scaling of SRE immediately gives the $q$ scaling of free energy
\be
I_q = {(q+1)^2\over 4q} I_1\ .
\ee

{\bf Three equal charges}~~~
Now we compute the holographic SRE from the STU with three equal charges,
\be
\el{kappa3}
\kappa_1=\kappa_2=\kappa_3=\kappa\ .
\ee
$T = T_0/q$ in this case gives
\be
\kappa = \frac{q-1}{2 q+1}\ .
\ee
The SRE and free energy can be obtained the same way as before
\be
{S_q\over S_1} = \frac{19 q^2+7q+1}{27 q^2}\ ,
\ee
\be
I_q = {(2q+1)^3\over 27q^2} I_1\ .
\ee

In fact the STU black hole with three equal charges can be regarded as the charged BPS solution \footnote{ The two different forms of metric are related to each other by a coordinate transformation (all
$Q_i$ are equal) $r^2 = {\hat r}^2 - Q_i$.}  (see \eg\cite{Buchel:2006gb}) in the five-dimensional ${\cal N}=2$ minimal supergravity theory, which can be obtained by further truncating the STU model. The bosonic part of five-dimensional minimal supergravity can be considered as an Einstein-Maxwell theory with a negative cosmological constant and also a Chern-Simons coupling. The charged BPS topological black hole solution for this theory is a natural extension of the four-dimensional one, the latter of which has been proposed as the dual of three-dimensional ${\cal N}=2$ superconformal field theories on branched three sphere $\mathbb{S}^3_q$ \cite{Huang:2014gca}.

{\bf Generic charges}~~~ One can also compute the holographic SRE from the black hole with three unequal charges. Note that all physical quantities can be regarded as functions only depending on $\hat \mu_i$, due to the generic relation between $\hat\mu_i$ and $\kappa_i$ (\ref{chemicalbulk}). To compare with the field theory result in the end, we first translate the chemical potentials (\ref{chemical4}) to the language of black hole 
\be
\hat\mu_1 = i(1-1/q)\frac{a}{3}\ ,\quad \hat\mu_2 = i(1-1/q)\frac{b}{3}\ ,\quad \hat\mu_3 = i(1-1/q)\lb 1-\frac{a+b}{3}\rb\ .
\ee With these input parameters, the holographic SRE can be obtained straightforwardly,
\be
{S_q\over S_1} = \frac{(a^2+ab-3a) (q-1) [3 q-b (q-1)]+3 q [b(b-3)(q-1)+9 q]}{27 q^2}\ .
\ee
Then the free energy was obtained, with $I_1 = -S_1$
\be
I_q=\frac{[a (q-1)-3 q]\, [b (q-1)-3 q]\, [a (q-1)+b (q-1)+3]}{27 q^2} I_1\ .
\ee Again, they agree with the localization result (\ref{f3_localization}) as well as the free field result (\ref{Sq3charges}) precisely.

\section{Conclusion and Discussions}\label{sec:conclusion}
In this work, we studied the four-dimensional superconformal field theories on sphere with conical singularity. We have mainly focused on ${\cal N}=4$ SYM theories on the branched sphere $\mathbb{S}^4_q$ with various background gauge fields. In the particular case of two U(1) background fields with equal values, we have shown that any ${\cal N}=2$ gauge theory can be embedded as the singular limit of the theory on a resolved sphere, whose partition function is essentially equal to that on an ellipsoid. We also wrote down the one-loop determinants for other backgrounds, which are crucial to determine the $q$-scaling behaviors of free energy in the large $N$ limit. 

For ${\cal N}=4$ SYM in each background, we computed the logarithmic term of free energy as well as supersymmetric R\'enyi entropy in the free field limit using heat kernel method. By carefully arranging the background fields as well as the R-charges of dynamical fields for ${\cal N}=4$ SYM, we showed that the $q$-dependence simply factorizes. Surprisingly, by evaluating the matrix integral coming from localization, we found the same $q$-dependence in the strong coupling regime, which implies that it is independent of coupling constant. \footnote{We note that \cite{Galante:2013wta} this is not the case for the non-supersymmetric R\'enyi entropy.} 

We found natural gravity duals of ${\cal N}=4$ SYM theory on $\mathbb{S}^4_q$ with various background gauge fields, the STU topological black holes. We thus provided the first concrete holographic dual of supersymmetric R\'enyi entropy in four dimensions which can be tested by field theory computation. We computed the holographic free energy and supersymmetric R\'enyi entropy from the black holes and found precise agreements with the corresponding large $N$ results of field theory. Based on these facts, we propose the TBH$_5$/qSCFT$_4$ correspondence, the higher dimensional extension of TBH$_4$/qSCFT$_3$ correspondence \cite{Huang:2014gca}. We believe further checks can be made for other observables, such as Wilson loops.

Notice that, currently we mostly restricted ourselves to the $\CN=4$ SYM on the $q$-deformed sphere. But the free field computations as well as the localization techniques are applicable for generic $\CN=2$ superconformal theories (SCFTs). It would be interesting to know whether the coupling independence of the $q$-scalings of supersymmetric R\'enyi entropies still holds in these cases. On the gravity side, however, it is rather unclear what the duals of SCFTs on the supersymmetric $q$-deformed sphere should be. Discussions about the gravity duals of ${\cal N}=2$ SCFTs 
that are relevant to this problem can be found in \cite{Kachru:1998ys,Lin:2004nb,Lin:2005nh,Gaiotto:2009gz,Gadde:2009dj,Rey:2010ry,Aharony:2012tz}.
 
Recently the ``universal behavior'' of R\'enyi entropy, related to the $q$-derivative, has been studied from different perspectives in \cite{Perlmutter:2013gua,Hung:2014npa, Lee:2014zaa}. We note that, similar investigations can be performed for the supersymmetric R\'enyi entropy $S_q$ we considered here. Naturally, the $q$-derivative of free energy is related to the correlator of stress tensor as well as R-current (see \cite{Closset:2012ru} for the discussion in three dimensions), because variation with respect to $q$ can be equivalently regarded as the variation with respect to both the metric component $g_{\tau\tau}$ and the background gauge field component $A_\tau$. 

In our case, the universal part of $S_q$ (also free energy) itself is independent of the coupling for ${\cal N}=4$ SYM, and therefore arbitrary $q$-derivatives of $S_q$ are independent of the coupling. It would be interesting to show explicitly how $S_q$ is protected by supersymmetry and receives no quantum corrections. We leave these questions for future works.


\section*{Acknowledgement}
We are grateful for useful discussions with Michael Duff, Kazuo Hosomichi, Hong Lu, Soo-Jong Rey and Tadashi Takayanagi. YZ would like to thank Kavli Institute for Theoretical Physics China, Asia Pacific Center for Theoretical Physics and Yukawa Institute for Theoretical Physics for their hospitality. The work of XH is supported by MOST Grant 103-2811-M-003-024.

\appendix
\section{Notations}
\label{app:notations}
On a 4-dimensional Euclidean spin manifold, the tangent space has the structure
group Spin$(4) \simeq SU(2) \times SU(2)$. Chiral spinors in two different Weyl
representations (chiral and anti-chiral) transform as doublets under the first and the second SU(2) respectively. Hence they are mutually independent. We use $\alpha, \beta=1,2$ indices for the first SU(2) and $\dot\alpha, \dot\beta=1,2$ indices for the second $SU(2)$. Indices are raised and lowered by the SU(2) invariant tensors
$\xe^{\xa\xb}$ and $\xe_{\xa\xb}$ (and also $\xe^{\dxa\dxb}$ and $\xe_{\dxa\dxb}$)
\be
\label{app:epsilon}
\xe^{12} = - \xe^{21} = -\xe_{12} = \xe_{21} = 1\,.
\ee
Note that the spinors are complex-valued, and a spinor $\xi_\alpha$ and its complex conjugate $\xi^*_\alpha$ transform under the same representation.

The gamma matrices $\xg^a$ (satisfying Clifford algebra $\xg^a\xg^b+\xg^b\xg^a = 2 \eta^{ab}$) reverse chirality and they can be written as
\be
\gamma^a= \left(
\begin{array}{cc}
 0 & \bar \xs^a \\
 \xs^a & 0 \\
\end{array}
\right)\, .
\ee 
The set of $2\times 2$ half-gamma matrices $(\sigma^a)_{\alpha\dot\alpha}$
($a=1,2,3,4$), together with $(\bar\sigma^a)^{\dot\xa \xa} = \xe^{\dot \xa\dot \xb} \xe^{\xa\xb} (\sigma^a)_{\xb\dot\xb}$, satisfy the algebra 
\[\sigma^a\bar \sigma^b+\sigma^b\bar \sigma^a = 2 \eta^{ab}\,.\]
In terms of Pauli matrices, they are given by
\be
\sigma^k = -i \tau^k\ , \quad \sigma^4=1\ ,\quad \bar\sigma^k = i \tau^k\ ,\quad \bar\sigma^4 = 1\ .~~(k=1,2,3)
\ee
The SU(2) transformation on spinors are generated by the self-dual tensor $\bar\sigma_{ab}$ ($\bar \sigma_{ab}=\frac12\varepsilon_{abcd}\bar \sigma^{cd}$) and the anti self-dual tensor $\sigma_{ab}$
\be \sigma_{ab}=\frac12(\sigma_a\bar\sigma_b-\sigma_b\bar\sigma_a),\quad \bar\sigma_{ab}=\frac12(\bar\sigma_a\sigma_b-\bar\sigma_b\sigma_a).\ee

Two Weyl spinors of opposite chiralities can combine to form a Dirac spinor (4-spinor)  $\zeta = (\xi_\xa, \bar \xi^{\dot \xa})$. In this paper, we mostly use the Language of $\CN=2$ supersymmetry. Capital letters $A,B$ are used to denote SU(2)$_R$ indices and they are raised and lowered by tensor $\xe^{AB}$
defined the same way as in \er{app:epsilon}. 

Following the usual convention, we use Greek letters for spacetime indices and Latin letters for internal indices.

\section{4-spinor to 2-spinor}
\label{app:4to2}
Here we rewrite the Killing spinor equations (\ref{4spinorkilling1})(\ref{4spinorkilling2})
in terms of 2-spinor notations. As we shall see, they are exactly matched with the Killing spinor equations (\ref{ks1}) in Section \ref{sec:Localization}.
First of all, $\zeta$ and $\zeta'$ can be decomposed as
\be
\zeta := \left(
\begin{array}{c}
\xi  \\
\bar\xi  \\
\end{array}
\right)\ ,\quad \zeta' := i\left(
\begin{array}{c}
\xi'  \\
\bar\xi'  \\
\end{array}
\right)\ ,
\ee and one obtains 
equations in terms of $2$-spinors. For (\ref{4spinorkilling1}), after decomposition we have
\ba
D_\mu\xi = -{i\over 2\ell}\sigma_\mu\bar\xi'\ ,\label{kse1}\\
D_\mu\bar\xi = -{i\over 2\ell}\bar\sigma_\mu\xi'\ ,\label{kse2}
\ea 
(\ref{4spinorkilling2}) can be equivalently written as additional constraints, by taking one more derivative up on (\ref{4spinorkilling1}). In terms of 2-spinors, they are
\ba
\sigma^\mu\bar\sigma^\nu D_\mu D_\nu \xi = M\, \xi\ ,\\
\bar\sigma^\mu\sigma^\nu D_\mu D_\nu \bar\xi = M\, \bar\xi\ .
\ea

\section{Resolved branched sphere and ellipsoid}
\label{App:Resolved}
\subsection{Resolved branched four-sphere}
The vielbein one-forms $E^a=E^a_\mu \rmd x^\mu$ for the resolved branched sphere
can be chosen as
\begin{equation}
 E^1=\sin\rho e^1,\quad
 E^2=\sin\rho e^2,\quad
 E^3=\sin\rho e^3+H \rmd \rho,\quad
 E^4=G \rmd \rho.
\end{equation}
where $e^a$ are vielbein of the three-dimensional ellipsoid in
polar coordinates $(\chi,\tau,\eta)$,
\begin{equation}
 e^1= \ell\cos\eta \rmd \chi,\quad
 e^2=q \ell \sin\eta \rmd \tau,\quad
 e^3=F \rmd \eta.
\end{equation}
$F,G,H$ in the metric components take the following forms
\ba
F(\eta, \rho ) & = & \sqrt{\frac{\cos ^2 \eta\, f_\xe(\sin \eta \sin \rho)^2}{1-\sin ^2\eta  \sin ^2\rho }+\frac{\ell^2\sin ^2 \eta \left(1-\sin ^2 \eta \sin ^2 \rho \right)}{\cos ^2 \rho \left(\cos ^2 \eta \tan ^2 \rho+1\right)^2}}\,, \nn
\el{metriccomp} H(\eta ,\rho) & = &  \frac{2 \sin 2 \eta  \cos \rho \left[f_\xe(\sin \eta  \sin \rho)^2-\ell^2\right]}{\left(2 \sin ^2\eta \cos 2 \rho+\cos 2 \eta+3\right) F(\eta,\rho)}\,,\\
G(\eta ,\rho)& = & \sqrt{\frac{\ell^2 f_\xe(\sin \eta \sin \rho)^2 \left(2 \sin ^2\eta  \cos 2 \rho +\cos 2 \eta +3\right)}{4 \left(\cos ^2\eta f_\xe(\sin \eta \sin \rho)^2+\ell^2\sin ^2 \eta \cos ^2\rho \right)}}\,.\nonumber
\ea
The components of the spin connection one-forms $\Omega^{ab}$ are given by,
\begin{eqnarray}
&&
 \Omega^{12}=0,\quad
 \Omega^{13}=-\frac \ell F\sin\eta \rmd \chi,\quad
 \Omega^{23}=\frac{q\ell}F\cos\eta \rmd \tau,\nonumber \\
&& \el{spinconS4}
 \Omega^{14}=\frac{\ell(\cos \eta \cos \rho\,  F+\sin \eta\, H)}{F G} \rmd \chi,\quad
 \Omega^{24}=\frac{q\ell (\sin \eta  \cos \rho\, F-\cos \eta \, H)}{F\, G} \rmd \tau,\\
&& \Omega^{34}=\frac{H \left(\sin \rho \p_\rho F-\p_\eta H \right)+\cos \rho  F\, H-G\,\p_\eta G}{\ell G}\rmd \eta \nn 
&& \qquad +\frac{ H \left[H\left(\sin \rho  \p_\rho F-\p_\eta H\right)+\cos \rho F\, H-G\, \p_\eta G\right]}{\ell \sin \rho F\, G}\rmd
\rho. \nonumber
\end{eqnarray}
Note that $\Omega^{12},\Omega^{13},\Omega^{23}$ are the spin connection
of the three-ellipsoid with vielbein $e^a$.

\subsection{Four-ellipsoid}
In a different set of coordinates (they are related to \er{Trans} by \er{coordtrf})
\begin{equation}
\begin{array}{rcl}
 x_0 &=& r\cos\rho, \\
 x_1 &=& \ell\sin\rho\,\cos\eta\,\cos\tau, \\
 x_2 &=& \ell\sin\rho\,\cos\eta\,\sin\tau, \\
 x_3 &=& \tilde\ell\sin\rho\,\sin\eta\,\cos\chi, \\
 x_4 &=& \tilde\ell\sin\rho\,\sin\eta\,\sin\chi,
\end{array}
\end{equation}
the metric of the four-ellipsoid \er{ellipsoids} becomes 
\be
\label{ellipsoid1}
\rmd s^2 = \sin^2 \rho(\tilde \ell^2\sin ^2 \eta\rmd
\tau^2+\ell^2 \cos^2\eta  \rmd\chi^2) + (f \sin \rho \rmd \eta + h \rmd \rho)^2
+ g^2 \rmd \rho^2\ ,
\ee
where $f,g,h$ are defined by
\begin{eqnarray}
\label{ellipfgh}
 f&:=& \sqrt{\ell^2\sin^2\eta+\tilde\ell^2\cos^2\eta},
 \nonumber \\
 g&:=& \sqrt{r^2\sin^2\rho+\ell^2\tilde\ell^2f^{-2}\cos^2\rho},
 \nonumber \\
 h&:=& \frac{\tilde\ell^2-\ell^2}f\cos\rho\sin\eta\cos\eta\,.
\end{eqnarray}

\section{5d BPS black hole}
\label{App:TBH}
The BPS black hole solution in the gauged theory was found in \cite{Behrndt:1998ns} and the solution is given by
\ba
\begin{aligned}
ds^2 = - {\CH^{-4/3}} f dt^2 + \CH^{2/3}
        \bigg(f^{-1} {dr^2} + 
        r^2 d\xS_{3,k} \bigg) \ ,\quad\quad\quad\quad \\ 
f = k + \frac {r^2}{L^2} \CH^2 \,, \quad \quad A_{t}^I = \sqrt{k} \lb \CH^{-1}Y^I\ -1\rb - \mu^I,\quad F_{r t}^I = \sqrt{k} \partial_r (\CH^{-1}Y^I), \\
X^I = \CH^{-\frac 1 3} Y^I\, , \quad \, \CH = \frac 1 6 C_{IJK} Y^I Y^J Y^K,\quad
\frac 1 2 C_{IJK} Y^J Y^K = H_I = 3V_I + \frac {Q_I}{r^2}\, , 
\label{BPSTBH}
\end{aligned}
\ea
where the last equation can be used to solve for $Y^I$. The parameter $k$ specifies the spatial curvature of $d\xS_{3,k}$. For flat space $\BR^3$ and three-sphere $\BS^3$, $k$ takes the values of $0$ and $+1$ respectively. For hyperbolic space $\BH^3$, $k=-1$. 
The explicit form of the metric for $d\xS_{3,k}$ can be chosen as
\be
d\xS_{3,k} = d\eta^2 + \Big({\sin \sqrt{k} \eta \over \sqrt{k}}\Big)^2
\Big(d\phi^2 + \sin^2\phi d\psi^2\Big) \ .
\ee

The supersymmetry transformation of gravitino reads 
\be
\el{KillingeqBH}
\delta\psi_\mu =\Big(\mathcal{D}_\mu + {\frac{ i}{8}}X_I (\xg_\mu{}%
^{\nu\rho} - 4 \delta_\mu{}^ \nu \xg^\rho) F_{\nu\rho}{}^I+{\frac{1}{2 L}}%
\xg_\mu X^IV_I-{\frac{3}{2 L}}i V_IA^I_\mu\Big) \epsilon,
\ee
where $\mathcal{D}_\mu$ is the covariant derivative and $\xg_{a\dots b}$ denotes the anti-symmetrized product of Gamma matrices with unit weight (\ie $\xg_{a b} = \frac 1 2 [\xg_a, \xg_b])$. 
$X_I$ is defined by 
\be X_I \equiv \frac 1 6 C_{IJK}X^J X^K = \frac 2 3 G_{IJ}X^J\,.\ee
The Killing spinor equation in this background is given by $\xd \psi_\mu = 0$. We choose the following f\"unfbein
\ba
\begin{aligned}
e^0 = \CH(r)^{-2/3}\sqrt{f}\rmd\tau,\quad e^1 = \frac{ {\CH(r)}^{1/3}}{\sqrt f}\rmd  r,\quad e^2 = r  {\CH(r)}^{1/3}\rmd \eta  \\
e^4 = \frac{r {\CH(r)}^{1/3} \sin \sqrt{k} \eta  }{\sqrt{k}} \rmd\phi  ,\quad \frac{r  {\CH(r)}^{1/3} \sin \phi \sin \sqrt{k} \eta  }{\sqrt{k}} \rmd\chi,
\end{aligned}
\ea
and the spin connection
\begin{align}
\begin{aligned}
\el{appeq:spinc}
\xo^{01}=\lb\frac{2 k \CH'(r)}{3 \CH(r)^2}+\frac{r^2 \CH'(r)}{3 L^2}+\frac{r \CH(r)}{L^2}\rb
\rmd\tau \,,\quad \xo^{02} = \xo^{03} = \xo^{04} = 0\\
\xo^{12} = -\frac{\left(r \CH'(r)+3 \CH(r)\right) {\sqrt f}}{3 \CH(r)} \rmd\eta \\
\xo^{13} = -\frac{\sin \sqrt{k}\eta   \left(r \CH'(r)+3 \CH(r)\right) {\sqrt f}}{3 \sqrt{k} \CH(r)}  \rmd\phi \\
\xo^{14} = -\frac{\sin\phi  \sin \sqrt{k}\eta   \left(r \CH'(r)+3 \CH(r)\right) {\sqrt f}}{3 \sqrt{k} \CH(r)} \rmd \psi\\ 
\xo^{23} = -\cos\sqrt{k}\eta\, \rmd \phi\,,\quad 
\xo^{24} = -\cos\sqrt{k}\eta \sin\phi\, \rmd \psi\,,\quad 
\xo^{34} = -\cos\phi\,\rmd \psi.
\end{aligned}
\end{align}
We can use the integrability condition $P \xe = 0$ to simplify \er{KillingeqBH}.
The projection operator $P$ is defined as
\begin{equation} \label{projector}
P :=\frac 1 2 + \frac 1 2 \Big(i x\Gamma _{0}+y \Gamma _{1}\Big) ,
\end{equation}
where 
\[
x=-\frac{\sqrt k}{\sqrt f}\ , \qquad y=-\frac{r}{\sqrt f L}\CH \ , \qquad
f = k + \frac{r^2}{L^2} \CH^2
\]
Note that we also have the following useful expressions,
\[V_I Y^I =  \CH + \frac 1 3 r \CH',\]
and ($0,1$ for internal indices)
\[X_I F_{01}^I = \frac{2\sqrt{k}} 3 \CH^{-\frac 4 3} \p_r \CH.\]
The temporal and spatial components of the Killing spinor equation are given by,  
\be
\begin{aligned}
\left( \partial _{t}-i\sqrt k g\right) \epsilon =0 \ , \\
\left( \partial _{r}-\frac{i\sqrt k}{2\sqrt f}\left( \frac{1}{r}+\frac{\CH^{\prime }} \CH
\right)\gamma_0 -%
\frac{1}{2}\left( \frac{1}{r}+\frac{\CH^{\prime }} {3\CH}\right)\right) \epsilon
=0, \\
\left( \partial _{\eta }+\frac{i\sqrt k}{2}\gamma _{012}\right) \epsilon =0 \ ,\\
\left( \partial _{\phi }+\frac{i}{2}\sin \sqrt k \eta \gamma _{013}-\frac{1}{2}%
\cos \sqrt k \eta \gamma _{23}\right) \epsilon =0, \\
\left(   \partial_\psi  +\frac{i}{2}\sin \sqrt k \eta\sin
\phi 
\gamma _{014}-\frac{1}{2}\cos\sqrt k \eta\sin
\phi \gamma _{24}-\frac{1}{2}\cos
\phi \gamma _{34}\right) \epsilon =0.
\end{aligned}
\ee
This type of equations can be solved \cite{Romans:1991nq}. We can solve the radial, temporal and angular equations separately. Time and angular components are solved first. The solution can be expressed as
\be
\epsilon  =e^{i \frac {\sqrt k} L t}e^{-{\frac{i}{2}}\gamma _{012}\sqrt k\eta }e^{+{\frac{1}{2}} \gamma _{23}\phi }e^{+{\frac{1}{2}}\gamma _{34}\psi }\varphi (r) 
\ee
The radial equation takes the form of
\[\p_r \xvp(r) = (a(r)+b(r)\xG_1)\xvp(r),\]
and $\xvp(r)$ also satisfies the constraint $P\xvp(r) = 0$ with $P$ in the form of
\[P = \frac 1 2 (1+ x(r)\xG_1+y(r)\xG_2)\ ,\]
where $\xG_{1,2}$ are matrices satisfying
\be \xG_1^2 = \xG_2^2 = 1, \quad \xG_1 \xG_2
+ \xG_2 \xG_1 = 0\ .\ee
Solution to this type of equation is provided in the appendix of \cite{Romans:1991nq}
\be \el{solspin}
\xvp(r) = (u(r)+v(r)\xG_2)\lb \frac {1-\xG_1} 2 \rb\xe_0,
\ee
where $u$, $v$ are defined by,
\be u = \sqrt{1 + x\over y} e^w,\quad v =- \sqrt{1 - x\over y} e^w,\quad
w(r) = \int^r a(r') dr',
\ee
and $\xe_0$ is an arbitrary constant spinor.
In our case, \er{solspin} gives
\be
\varphi (r) = {1\over2}\sqrt{\frac L r }
{\cal V}^{-{1\over2}}\Big(\sqrt{\sqrt f+k}-\sqrt{\sqrt f-k}\gamma_1\Big) e^{{1\over2}\int^{r}
d\bar r \Big({1\over \bar r}+{1\over 3}{{\cal V}'\over{\cal V}}  \Big)} 
\; (1-i\gamma_0)\epsilon_0 \,.
\ee

\section{Multiple Gamma Function}
\label{MGamma}
Multiple Gamma function $\xG_n(x,\vec \xo)$ ($\vec \xo$ being an $n$-vector
$(\xo_1, \dots, \xo_n)$)
is defined by
\be
\xG_n(x,\vec \xo):=\prod_{m_1 \dots m_n = 0}^\infty (m_1 \xo_1 + \dots m_n
\xo_n + x)^{-1}.
\ee

The function $\log \xG_n(x,\vec \xo)$ can be expanded in the large $|x|$ limit \cite{Alday:2014bta} (see also \cite{Ruijsenaars:2000})
\ba
\log \xG_n(x,\vec \xo)  & = & \frac{(-1)^{n+1}}{n!} B_{n,n}(x) \log x +(-1)^{n} \sum_{k=0}^{n-1} \frac{B_{n,k}(0)x^{n-k}}{k! (n-k)!}\sum_{\ell=1}^{n-k} \frac{1}{\ell}   +\mathcal{O} \left( w^{-1} \right),\;\qquad \label{logGexpansion}
\ea
where the functions $B_{n,m}\left( x \right)$ are the so-called multiple Bernoulli polynomials. For $n=2, \vec \xo = (a,b)$, \er{logGexpansion} reduces
to
\ba
\log \xG_2(x,(a,b))& =& -\frac{x^2 \log x}{2 a b}+\frac{3 x^2}{4 a b}+\frac{1}{2}  \left(\frac{1}{a}+\frac{1}{b}\right) x\log x -\frac{1}{2}  \left(\frac{1}{a}+\frac{1}{b}\right)x\nn
& &-\left[\frac{1}{12} \left(\frac{a}{b}+\frac{b}{a}\right)+\frac{1}{4}\right] \log x+\CO(\iv x).\el{app:G2lartexexp}
\ea
Similarly for $n=3, \vec \xo = (a,b,c)$, we have
\ba
\log \xG_3(x,(a,b,c))& =& \frac{1}{3!}\log x \left[\frac{1}{a b c}x^3-\frac{ 3 (a+b+c)}{2 a b c}x^2+\frac{ (a+b+c)^2+a b+a c+b c}{6 a b c}x\rc \nn 
& & \lc-\frac{(a+b+c)(a b+a c+b c)}{4 a b c}\right]- \frac 1 {3!} \lsb \frac{11}{6a b c}x^3 -\frac{ 9 (a+b+c)}{4 a b c}x^2\rc\nn
& & \lc +\frac{ (a+b+c)^2+a b+a c+b c}{2 a b c}x \rsb+\CO(\iv x). \el{G3lartexexp}\ea

\end{document}